\documentclass[runningheads]{llncs2e/llncs}
\makeatletter
\renewcommand*\l@author[2]{}
\renewcommand*\l@title[2]{}
\makeatletter
\usepackage[noadjust,nosort]{cite}
\usepackage[bookmarks=true,colorlinks=true]{hyperref}
\usepackage{xspace}
\usepackage{figlatex,wrapfig}
\usepackage{listings,amssymb,mathtools}
\usepackage{amsmath}
\usepackage{array,multirow}
\usepackage{tikz}
\usetikzlibrary{positioning}
\usetikzlibrary{shapes}
\usetikzlibrary{shapes,fit,backgrounds}
\usetikzlibrary{decorations.pathreplacing}
\usetikzlibrary{calc}
\usepackage{enumerate}
\usepackage{galois}
\usepackage{proof}
\usepackage{trfrac/trfrac}
\usepackage{subfig}
\usepackage{threeparttable}

\usepackage[ruled,vlined]{algorithm2e}
\usepackage{algorithmic}

\usepackage{etoolbox}

\usepackage[normalem]{ulem}

\makeatletter
\newcommand\footnoteref[1]{\protected@xdef\@thefnmark{\ref{#1}}\@footnotemark}
\makeatother

\makeatletter
\@ifpackageloaded{stix}{%
}{%
\DeclareFontEncoding{LS2}{}{\noaccents@}
\DeclareFontSubstitution{LS2}{stix}{m}{n}
\DeclareSymbolFont{stix@largesymbols}{LS2}{stixex}{m}{n}
\SetSymbolFont{stix@largesymbols}{bold}{LS2}{stixex}{b}{n}
\DeclareMathDelimiter{\lBrace}{\mathopen} {stix@largesymbols}{"E8}%
{stix@largesymbols}{"0E}
\DeclareMathDelimiter{\rBrace}{\mathclose}{stix@largesymbols}{"E9}%
{stix@largesymbols}{"0F}
}
\makeatother

\definecolor{brown}{rgb}{0.59, 0.29, 0.0}
\definecolor{green}{rgb}{0.0, 0.5, 0.0}
\definecolor{orange}{rgb}{1.0, 0.62, 0.0}
\definecolor{purple}{rgb}{0.5, 0.0, 0.5}
\definecolor{red}{rgb}{1.0, 0.0, 0.0}
\definecolor{blue}{rgb}{0.0, 0.0, 1.0}

\newcommand {\ie}{{\em i.e.},\xspace}
\newcommand {\eg}{{\em e.g.},\xspace}
\newcommand{\defeq}{\stackrel{\mathclap{{\tiny \mbox{def}}}}{=}\ }
\newcommand{\andspace}{\mathbin{\wedge}}
\newcommand{\orspace}{\mathbin{\vee}}
\newcommand{\intersectionof}[2]{\bigcap\limits_{#1}^{#2}}
\newcommand{\unionof}[2]{\bigcup\limits_{#1}^{#2}}

\newcommand{\tuple}[1]{\textcolor{orange}{(}#1\textcolor{orange}{)}}
\newcommand{\progstate}[2]{(\tuple{#1},\tuple{#2})}

\newcommand{\seqb}[2]{{#1} {\color{brown}\rightarrow^{sb}} {#2} }

\newcommand{\rf}[2]{{#1} {\color{blue}\rightarrow^{rf}} {#2} }

\newcommand{\setRF}{{\color{blue}\mathtt{rf}}}

\newcommand{\proofrule}[3]{$ \trfrac[\textsc{#1}] {#2} {#3}$ }

\newcommand{\vars}{\mathcal{V}}
\newcommand{\loc}{\mathcal{L}}

\newcommand{\lab}{\ell}

\newcommand{\st}{\mathtt{st}}
\newcommand{\ld}{\mathtt{ld}}
\newcommand{\rmw}{\mathtt{rmw}}
\newcommand{\lock}{\mathtt{lock}}
\newcommand{\unlock}{\mathtt{unlock}}
\newcommand{\store}[2]{\st\ #1\ #2}
\newcommand{\load}[1]{\ld\ #1}
\newcommand{\rmwinst}[3]{\rmw\ #1\ #2\ #3}

\newcommand{\lockinst}[1]{\lock\ #1}
\newcommand{\unlockinst}[1]{\unlock\ #1}
\newcommand{\tool}{\texttt{PRIORI}\xspace}
\newcommand{\conla}{\mathcal{T}\xspace}
\newcommand\sbullet[1][.5]{\mathbin{\vcenter{\hbox{\scalebox{#1}{$\bullet$}}}}}
\newcommand{\TO}{\textcolor{red}{TO}}
\newcommand{\red}[1]{\textcolor{red}{#1}}
\newcommand\conleq{\mathrel{\ooalign{$\subseteq$\cr
  \hidewidth\raise.15ex\hbox{$\sbullet\mkern2mu$}\cr}}}
\newcommand{\mo}[1]{\mathrel{\leqslant_{#1}}\xspace}
\newcommand{\inmo}[3]{#2 \mathbin{\mo{#1}} #3\xspace}

\newcommand{\totordset}[2]{\lBrace \leqslant \rBrace_{#1}}
\newcommand{\modord}[2]{#1_{#2}\xspace}
\newcommand{\pomo}[1]{\mathrel{\preccurlyeq_{#1}}\xspace}
\newcommand{\inpomo}[3]{#2 \mathbin{\pomo{#1}} #3\xspace}

\newcommand{\append}{\mathbin{\Diamond}}
\newcommand{\iscons}{\mathbin{\uparrow}}

\newcommand{\isncons}{\mathbin{\nuparrow}}
\newcommand{\isext}{\mathbin{\lhd}}
\newcommand{\isnext}{\mathbin{\ntriangleleft}}
\newcommand{\join}{\mathbin{\sqcup}}
\newcommand{\meet}{\mathbin{\sqcap}}
\newcommand{\widen}{\mathbin{\nabla}}
\newcommand{\leqlattice}{\mathbin{\sqsubseteq}}
\newcommand\squplus{\mathbin{\ooalign{$\sqcup$\cr%
			\hfil\raise0.42ex\hbox{$\scriptscriptstyle+$}\hfil\cr}}}

\newcommand{\abs}[1]{{#1}^\sharp}

\newcommand{\reason}[1]{\mbox{(#1)}}

\newcommand{\xmark}{$\times$}
\newcommand{\assign}{\coloneqq}

\newtheorem{dfn}{Definition}
\newcommand*\circled[1]{\tikz[baseline=(char.base)]{
    \node[shape=circle,draw,inner sep=2pt,fill=brown!20,scale=0.65] (char) {#1};}}

\newcommand{\duet}{{\small \sc DUET}\ }
\newcommand{\TMAI}{{\small TMAI}\xspace }
\newcommand{\TSO}{{\small TSO}\ }
\newcommand{\PSO}{{\small PSO}\ }
\newcommand{\SC}{{\small SC}\ }
\newcommand{\RMO}{{\small RMO}\ }
\newcommand{\POET}{{\small POET}\ }
\newcommand{\PO}{{\small PO}\xspace}
\newcommand{\RA}{{\small RA}\xspace }
\newcommand{\vbmc}{{\small \sc VBMC}\xspace}
\newcommand{\rcmc}{{\small \sc RCMC}\xspace}

\begin{document}

\title{Thread-modular Analysis of Release-Acquire Concurrency}

\author{
Divyanjali Sharma \inst{1} \and\
Subodh Sharma \inst{1}}
\institute{Indian Institute of Technology Delhi
\email{\{divyanjali,svs\}@cse.iitd.ac.in}}
\maketitle 
\begin{abstract}
%
%
We present a thread-modular abstract interpretation (\TMAI) technique 
to verify programs under the {\em release-acquire} (\RA) memory model
for safety property violations.
The main contributions of our work are: we capture the
execution order of program statements as an abstract domain, and 
propose a sound {\em upper approximation} over this domain to efficiently reason
over \RA concurrency. The proposed domain is general in its
application and captures the ordering relations as a first-class
feature in the abstract interpretation theory. In particular, the domain
represents a set of sequences of {\em modifications} of a global
variable in concurrent programs as a {\em partially ordered} set.
%
Under the upper approximation, older {\em sequenced-before} stores of a
global variable are forgotten and only the latest stores per variable
are preserved.
We establish the soundness of our proposed abstractions 
%
and implement 
them 
in a prototype abstract interpreter
called \tool.
The evaluations of \tool on existing and challenging \RA benchmarks
demonstrate that the proposed technique is not only competitive in
refutation, but also in verification.  \tool shows significantly fast
analysis runtimes with higher precision compared to recent
state-of-the-art tools  
for \RA concurrency.
\end{abstract}

\section{Introduction} \label{sec:intro}
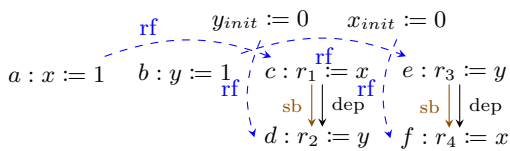
\begin{wrapfigure}{l}{0.6\linewidth}
\centering
\tikzset{every picture/.style={line width=0.75pt}} 
\begin{tikzpicture}[x=1em,y=1em,yscale=1,xscale=1]
\tikzstyle{every node}=[font=\small]
%

\node (inity)  {$ y_{init} \assign 0 $};
\node (initx) [right =7pt of inity] {$ x_{init} \assign 0 $};
\node (wy1) [below left=5pt and -15pt of inity] {$ b: y \assign 1 $};
\node (rx1) [right =5pt of wy1] {$ c: r_1 \assign x $};
\node (wx1) [left =5pt of wy1] {$ a: x \assign 1 $};
\node (ry0) [below =10pt of rx1] {$ d: r_2 \assign y $};
\node (ry1) [right =5pt of rx1] {$ e: r_3 \assign y $};
\node (rx0) [below =10pt of ry1] {$ f: r_4 \assign x $};

\draw [dashed,->,>=stealth,color=blue,thin] (initx.south) to[bend right=20] node[midway,left]{rf} (rx0.west);
\draw [dashed,->,>=stealth,color=blue,thin] (inity.south) to[bend right=20]  node[midway,left] {  rf} (ry0.west); %
\draw [dashed,->,>=stealth,color=blue,thin] (wx1) to[bend left=20] node[near start,above] {  rf} (rx1);
\draw [dashed,->,>=stealth,color=blue,thin] (wy1) to[bend left=20] node[midway,below] { rf} (ry1);

\draw [solid,->,>=stealth,color=brown,thin] ($ (rx1)+(-2pt,-4pt) $) -- node[midway,left] {\scriptsize sb} ($ (ry0)+(-2pt,4pt)$);
\draw [solid,->,>=stealth,color=brown,thin] ($ (ry1)+(-2pt,-4pt) $) -- node[midway,left] {\scriptsize sb} ($ (rx0)+(-2pt,4pt) $);

\draw [->,>=stealth,thin] ($ (rx1)+(2pt,-4pt) $) -- node[midway,right] {\scriptsize dep} ($ (ry0)+(2pt,4pt) $);
\draw [->,>=stealth,thin] ($ (ry1)+(2pt,-4pt) $) -- node[midway,right] {\scriptsize dep} ($ (rx0)+(2pt,4pt) $);

\node (property) [below left=3pt and -90pt of ry0]{$P: r_1=1 \andspace r_3=1 \implies r_4 \neq 0 \andspace r_2 \neq 0$ };
\node (name) [below =3pt of property]{\texttt{(IRIW)}};
\end{tikzpicture}
\caption{IRIW execution graph  with {\em reads-from} (rf) and
{\em sequence-before} (sb) edges}
\label{fig:iriw-rel}	
\vspace{-1em}
\end{wrapfigure}

We investigate the problem of verifying programs with assertions
executing under the {\em release-acquire} (\RA) fragment of the C11
standard~\cite{C11} where every store is a {\em release} write and
every load is an {\em acquire} read. 
%
The reachability problem under the \RA model (with {\em compare-and-swap})
has been recently shown to be undecidable \cite{AbdullaKrishna-PLDI19}. The 
model is described  {\em axiomatically}
and correctness of programs under the model is defined by acyclicity
axioms, which can appear obscure.
Notwithstanding the undecidability result, \RA model is still one of
the cleaner subsets of the C11 standard with relatively
well-behaved semantics
and has been a subject of active study in recent times
\cite{Batty-POPL11,RCU-PLDI15,LahavVafeiadis-POPL16,KangVafeiadis-POPL17,KokologiannakisVafeiadid-POPL18,AbdullaKrishna-PLDI19}.
An incomplete but intuitive understanding of \RA concurrency is
usually provided through reorderings --  
the redordering of an acquire load
(or release store)  with any access that follow
(or precede) it in program order is disallowed.
The \RA model indeed provides weaker guarantees than {\small SC}, which
allows for the construction of high performance implementations (\eg
read-copy-update synchronisation~\cite{RCU-PLDI15}) without making
programmability overly complex.
However, as noted in \cite{LahavVafeiadis-POPL16}, \RA programs can
produce counter-intuitive outcomes that are unexplainable via
interleaving of instructions. 
Consider the example {\em execution graph} (or just execution) of
a 4-threaded program (IRIW) in Figure~\ref{fig:iriw-rel}.  It shows
through appropriate {\em reads-from} (rf), {\em sequence-before}
(sb) and data/control {\em dependency} (dep)
edges that the property $P$ can evaluate to false under \RA model
(\ie $r1=r3=1, \ r2=r4=0$). However, when the execution is interpreted
under interleaving execution semantics (such as in {\small SC, TSO},
and {\small PSO}), the property is evidently valid because of a
total ordering between $a$ and $b$ (\ie $a$ before $b$ or vice-versa).
Nonetheless, there are some fascinating aspects of \RA semantics --
(i) a total order on the stores of each global memory location
(called the {\em modification order}) that restricts loads 
reading from overwritten stores, and (ii)  when a load instruction of a thread
$t$ observes (or {\em synchronizes} with) a store from another thread
$t'$, then all the prior stores observed by $t'$ up to the
synchronizing store also become observable to $t$.
It is worth noting that this lack of immediate global visibility of
updates, as mentioned in (ii) above, makes programs under \RA semantics
naturally amenable to {\em localized} or {\em thread-modular}
reasoning, which is a well-considered area of research.
\newline
\indent Thread-modular approaches are known to be sound for safety
properties~\cite{Henzinger-CAV03}.
The basic idea behind thread-modular reasoning is
to verify each thread separately with an environment assumption to
model the effects of the execution of other threads. The environment
assumption of each thread is usually specified by a relation (referred
to as {\em interference relation} in this paper), which includes all
the stores to global memory of other threads that may affect the loads
of the thread. The analysis proceeds iteratively until for each
thread the effects of its environment assumption on its operations
reach a fix-point.
%
%
As a model checking approach, they were first introduced for programs
under \SC semantics~\cite{Flanagan-SPIN03}.
In the recent past, several thread-modular contributions
\cite{SuzanneMine-APLAS18,Mine-VIMCAI17,SuzanneMine-SAS16,KusanoWang-FSE16,KusanoWang-FSE17}
have been presented in the context of verifying programs under weak
memory models such as {\small TSO, PSO} and {\small RMO}.
%
However, in our observation,  prior proposals
run into fundamental limitations when applying them to \RA or other
{\em non-multicopy-atomic} memory models such as {\small ARMv7} and {\small POWER}.
\newline
%
%
\indent Techniques presented in~\cite{SuzanneMine-SAS16,SuzanneMine-APLAS18}
model store buffers to analyze {\small TSO} and {\small PSO} programs.
Evidently, \RA program behaviors cannot be simulated using
store buffers \cite{LahavVafeiadis-POPL16}. Thus, extending these
contributions is not
feasible without re-modeling for the operational semantics of
\RA~\cite{KangVafeiadis-POPL17}.
Contributions such as \textsc{Watts}~\cite{KusanoWang-FSE16} and
\textsc{FruitTree}~\cite{KusanoWang-FSE17}
extend \TMAI with lightweight flow- and context-sensitivity. However,
they do not capture inter-thread ordering dependencies
beyond two threads.
%
Finally, the abstract interpretation technique used in
\duet \cite{farzan_duet_2013}
is neither
thread-modular nor geared for \RA programs.
While \duet performs analysis with an unbounded number
of threads, it may infer 
gross over-approximations on some simple programs. 
Consider the following program where initially  $x = 0$ :
\begin{tabular}{c||c}
$a: x++$ & $b: x++$. 
\end{tabular}
 \duet will infer the value $x = \infty$
at termination.
{\sc FruitTree} \cite{KusanoWang-FSE17} also
suffers from the same imprecision, though it does not terminate.
\noindent{\bf Contributions and Outline}:
%
In this paper, (C1) as our first contribution, we propose a \TMAI technique
(see \textsection\ref{sec:thr-mod}) for \RA programs using a novel
abstract domain which is based on partial orders (\PO). 
%
%
%
The proposed domain succinctly captures
abstract ordering dependencies among instructions in a program
(see \textsection\ref{sec:abs-sem}). While the use of partial orders
to analyze concurrency is well-known, to the best of our knowledge
this is the first work that formulates the ordering information
as an abstract domain.
In particular, we model the concrete program
semantics as a set of total orders on stores per global variable, also
known as {\em modification order} ({\tt
mo})(see \textsection\ref{sec:con-sem}). A collection of
{\tt mo}s are then represented as a \PO domain.
Notably, the use of \PO domain has
the following merits: (M1) \PO domain is general in its scope and is
applicable beyond \RA concurrency (see \textsection\ref{sec:mo-po-lat}
and \textsection\ref{subsec:generic-mm}).  (M2) Introduction of
ordering information as a first-class object in abstract
interpretation theory permits further abstractions or refinements on
the object, an instance of which is 
presented in contribution (C2).
%

%
\hypertarget{c2}{(C2)} We present 
an {\em abstract upper approximation} of \PO domain
(see \textsection\ref{subsec:abs-po-mo}) where
only 
the {\em latest stores} per thread per variable are preserved and 
all the older $\color{brown}{sb}$-ordered stores are forgotten. 
%

(C3) Furthermore, to establish that our analysis preserves soundness and is
terminating, we show that (i) the lattice corresponding to the
abstract semantics is {\em complete}, (ii) establish
a  {\em Galois connection} between the concrete and \PO
domains, (iii) prove that the {\em abstract upper approximation} is
sound, and (iv) provide a widening operator
to ensure termination of the analysis

(C4) Finally, we implement our proposal in a prototype tool
called \tool, and demonstrate its effectiveness in refutation and
verification of \RA programs by comparison with
recent state-of-the-art tools in the \RA domain (see \textsection\ref{sec:experiments}).  
%

%

We present related work in \textsection\ref{sec:related-work} followed
by an intuitive account of our contributions with the help of
examples in \textsection\ref{sec:ovw}.

\section{Related Work} \label{sec:related-work}
Weak memory models, in particular C11 model, have been topics
of active research in recent years. Many studies have provided
proof and logic frameworks
\cite{VafeiadisChinmay-OOPSLA13,TuronVafeiadis-OOPSLA14,LahavVafeiadis-ICALP15,DokoVafieadis-VMCAI16}
and recommended
strengthening the
C11 models \cite{LahavVafeiadis-POPL16,KangVafeiadis-POPL17}. Many
existing contributions have proposed stateless
model checking algorithms for \RA programs using state-reduction
techniques such as dynamic partial order reduction or event
structures \cite{Kokologiannakis19,Abdulla-OOPSLA2018,KokologiannakisVafeiadid-POPL18,Norris-TPLAS2016,LahavVafeiadis-PLDI17,ZhangKW15}.  
\newline
\indent In contrast, there have been relatively fewer investigations of \RA
concurrency using symbolic analysis.  While some works have
explored using \TMAI (which have already been discussed
in \textsection\ref{sec:intro}), others have proposed BMC as solutions to
verify programs under models such as \TSO, \PSO and \RMO.
%
%
%
%
\newline
\noindent{\bf Bounded Model Checking.} BMC contributions 
in~\cite{Gavrilenko-CAV19,AbdullaKrishna-PLDI19,AlglaveKT13} operate
by placing a bound on the number of loop
unrollings or on the number of contexts or both.
{\em Dartagnan}~\cite{Gavrilenko-CAV19} is a BMC framework that
offers support for parameterized reasoning over memory
models.
While, in principle, 
Dartagnan can perform bounded reasoning of \RA programs, it currently does
not support \RA semantics.
\newline
%
\indent VBMC~\cite{AbdullaKrishna-PLDI19}, a recent BMC solution for \RA concurrency, 
works with an additional bound
on the number of {\em views} in a \RA program -- a \emph{view} of a
thread is a collection of timestamps of the latest stores it has
observed for each variable. A \emph{view-switch} takes place when a
load operation in a thread, say $ t_2 $, reads from a store in a
thread, $ t_1 $, with a timestamp higher than that of any variable in
the view of $ t_2 $.  While efficient in refutation,
VBMC fails to discover property violations in programs which are parametric
in the number of readers where the number
of view-switches required is beyond the default bound of two
(see Appendix~\ref{appendix:rel-work-details} for a detailed discussion).
\newline
\noindent{\bf \PO encodings and unfoldings.}
The use of partial order encodings is diverse and rich in areas of
concurrent program verification and testing. The works in \cite{WangKGG09,Huang016,ForejtJKNS17}
use partial order encodings in dynamic verification tools
to predictively reason about  multithreaded and message-passing programs.
Partial order encoding presented in 
\cite{AlglaveKT13} relies on the axiomatic semantics of memory models such as \SC, 
Intel X86 and {\small IBM} {\small POWER} and is implemented in a BMC
tool. The contributions in \cite{RodriguezSSK15} and \cite{heljanko15}
use unfolding semantics to verify and test \SC programs, respectively.
\newline
\indent A recent study (\POET \cite{SousaRDK17}) combines unfolding semantics with abstract
interpretation.  The solution they have proposed is elegant and close
to our proposal, but with several fundamental differences: (D1) \POET
defines the unfolding under a variant of the {\em independence}
relation used in the partial order reduction theory~\cite{modelChecking}.
Evidently, the independence relation assumes an interleaving
model of computation. While unfoldings can capture {\em true
concurrency}, the independence relation fundamentally limits their
general applicability and restricts \POET's application to only those
memory models that can be explained with interleavings. As a result, we
have found \POET's technique to be unsound for \RA programs. (D2) \POET
uses unfoldings as an auxiliary object which is external to the
abstract interpretation theory. Thus, it is not straightforward to
define further abstractions on the unfolding object once created. On
the contrary, in our proposal, the \PO domain is treated as a
first-class object of the abstract interpretation theory, which is
open to further abstractions as is witnessed in our
contribution \hyperlink{c2}{(C2)}. (D3) \POET is not thread-modular and 
navigates an  unfolding object of an entire program which is much larger
than the \PO domains maintained per location per variable in our technique.

\section{Overview} \label{sec:ovw}
We provide an overview of thread-modular analysis using PO domain with
the help of small examples.

Let $a$ and $b$ be load and store operations, respectively from
different threads to a global memory location. The store $b$
is then called an {\em interference} for load $a$ (denoted by $\rf{a}{b}$, since
$b$ can potentially read from $a$).

\subsection{Thread Modular Analysis with Partial Order Domain}
\label{subsec:tmai-po-overrview}
Consider the message passing program ({\tt MP}) shown
below on the left. Under \RA semantics
if $r_1 = 1$, then $r_2 = 0$ is infeasible. Thus,
property $P$ is known to be valid.

\begin{minipage}{0.35\textwidth}
	\begin{tabular}{c||c}
		\multicolumn{2}{c}{\tt (MP)} \\
		$a: x \assign 1$ &$c: r_1 \assign y$  \\
		$b: y \assign 1$ &$d: r_2 \assign x$  \\
		\multicolumn{2}{c} {$P: r_1=1 \implies r_2=1$} 
	\end{tabular}
\end{minipage}
\hspace{1em}
\begin{minipage}{0.65\textwidth}
	\tikzset{every picture/.style={line width=0.75pt}} 
\begin{tikzpicture}[x=1em,y=1em,yscale=1,xscale=1]
\tikzstyle{every node}=[font=\footnotesize]
\node (initx) {$ x_{init} \assign 0 $};
\node (inity) [right =6pt of initx] {$ y_{init} \assign 0 $};
\node (wx1) [below =5pt of initx] {$ a: x \assign 1 $};
\node (wy1) [below =8pt of wx1]   {$ b: y \assign 1 $};
\node (ry1) [below =5pt of inity] {$ c: r_1 \assign y $};
\node (rx1) [below =8pt of ry1] {$ d: r_2 \assign x $};
\node (sigA) [left=30pt of wx1] {$ (\ $};
\node (poaXn1) [circle,fill=black,inner sep=0pt,minimum size=3pt, right=4pt of sigA] {};
\node (poaXa) [left=1pt of poaXn1, inner sep=1pt] {$a$};
\node (poaXfit) [draw,fit=(poaXn1)(poaXa), color=orange, thin, inner sep=1pt] {};
\node (poaY) [below right =-5pt and -2pt of poaXfit] { , };
\node (poaYempty) [right=6pt of poaXfit] {};
\node (poaYfit) [draw,fit=(poaYempty), color=orange, thin, inner sep=1pt] {};
\node (valA) [right =-2pt of poaYfit] {)};
\node (poA) [fit=(poaXfit)(poaYfit)(valA)] {};

\node (sigB) [left=30pt of wy1] {$ (\ $};
\node (pobXn1) [circle,fill=black,inner sep=0pt,minimum size=3pt, right=4pt of sigB] {};
\node (pobXa) [left=1pt of pobXn1, inner sep=1pt] {$a$};
\node (pobXfit) [draw,fit=(pobXn1)(pobXa), color=orange, thin, inner sep=1pt] {};
\node (pobX) [below right =-5pt and -2pt of pobXfit] { , };
\node (pobYn1) [circle,fill=black,inner sep=0pt,minimum size=3pt, right=12pt of pobXfit] {};
\node (pobYb) [left=1pt of pobYn1, inner sep=0pt] {$b$};
\node (pobYfit) [draw,fit=(pobYn1)(pobYb), color=orange, thin, inner sep=1pt] {};
\node (valB) [right =-2pt of pobYfit] {)};
\node (poB) [fit=(pobXfit)(pobYfit)(valB)] {};

\node (sigC) [right=-5pt of ry1] {$ (\ $};
\node (pocXn1) [circle,fill=black,inner sep=0pt,minimum size=3pt, right=4pt of sigC] {};
\node (pocXa) [left=1pt of pocXn1, inner sep=1pt] {$a$};
\node (pocXfit) [draw,fit=(pocXn1)(pocXa), color=orange, thin, inner sep=1pt] {};
\node (pocX) [below right =-5pt and -2pt of pocXfit] { , };
\node (pocYn1) [circle,fill=black,inner sep=0pt,minimum size=3pt, right=12pt of pocXfit] {};
\node (pocYb) [left=1pt of pocYn1, inner sep=0pt] {$b$};
\node (pocYfit) [draw,fit=(pocYn1)(pocYb), color=orange, thin, inner sep=1pt] {};
\node (valC) [right =-2pt of pocYfit] {)};
\node (poC) [fit=(pocXfit)(pocYfit)(valC)] {};

\node (sigD) [right=-5pt of rx1] {$(\ $};
\node (podXn1) [circle,fill=black,inner sep=0pt,minimum size=3pt, right=4pt of sigD] {};
\node (podXa) [left=1pt of podXn1, inner sep=1pt] {$a$};
\node (podXfit) [draw,fit=(podXn1)(podXa), color=orange, thin, inner sep=1pt] {};
\node (podX) [below right =-5pt and -2pt of podXfit] { , };
\node (podYn1) [circle,fill=black,inner sep=0pt,minimum size=3pt, right=12pt of podXfit] {};
\node (podYb) [left=1pt of podYn1, inner sep=0pt] {$b$};
\node (podYfit) [draw,fit=(podYn1)(podYb), color=orange, thin, inner sep=1pt] {};
\node (valD) [right =-2pt of podYfit] {)};
\node (poD) [fit=(podXfit)(podYfit)(valD)] {};


\draw [dashed,->,>=stealth,color=blue,thin] (wy1) -- node[midway,above] {rf} (ry1);

\draw [solid,->,>=stealth,color=brown,thin] (wx1) -- node[midway,left] {sb} (wy1);
\draw [solid,->,>=stealth,color=brown,thin] (ry1) -- node[midway,right] {sb} (rx1);
%
\draw [dashed,->,>=stealth,color=orange,thin] (wx1) -- node[midway,below] {hb} (rx1);

\end{tikzpicture}
\end{minipage} 
%
%
\noindent {\bf Program state.}  Let
poset $ PO_x$ and $ V_x$ represent the partial order on the observed
stores and the abstract value of variable $x \in \vars$ where $\vars$
is the set of all shared variables in a program. We present the
program state (or just state) at each program location (or just
location) as a tuple $(\Pi_{x \in \mathcal{V}} PO_x
, \Pi_{x \in \mathcal{V}} V_x)$, where $\Pi$ is a cartesian product
operator defined over indexed family of sets. Consider an
execution of {\tt (MP)} shown above on the right. At location $a$, the
state in components is: 
$PO_x = (\{a\}, \emptyset), PO_y = (\emptyset, \emptyset), V_x = \{1\},
V_y = \emptyset$ (Note that the second argument of a poset is the
ordering relation). For brevity, we only show the posets of variables
(as location-labeled Hasse diagram in a box)
and suppress the abstract value in the above and future illustrations.

\noindent {\bf Interferences}. 
Consider the above {\tt MP}
example again. Thread 1 has no loads; therefore, has no computable interferences.
In thread 2, the set of
interferences at locations $c$ and $d$ are $ \{\rf{b}{c}, \rf{\mathtt{ctx}}{c}\}
$ and $ \{\rf{a}{d}, \rf{\mathtt{ctx}}{d}\} $, respectively.
Note that $\mathtt{ctx}$ refers to a special label representing {\em context} --
\ie in the absence of any                          
interfering stores, a load instruction will either read from the
latest preceding {\tt po} (program order) store or from the store values
that have traveled embedded in the program states up to that load
instruction.

\noindent{\bf TMAI.}
In the first iteration, the states of thread 1 are computed as shown
in the above illustration for locations $a$ and $b$. In thread 2, in
the absence of any interefering store, the states are computed with
the information from $\mathtt{ctx}$, where $ PO_x $ and $ PO_y $ are
empty.  Therefore, both at $c$ and $d$ we have : $PO_x =
(\emptyset, \emptyset), PO_y = (\emptyset, \emptyset)$.

In the second iteration, 
 the interference $ \rf{b}{c} $ is applied, and the effects of
{\em all} the instructions 
prior to $b$ from thread 1 are carried to $c$ and $d$.
Thus, at $c$, we have: $PO_x =                                 
(\{\mathtt{a}\}, \emptyset), PO_y = (\{\mathtt{b} \}, \emptyset)$.
As a result, the effect of $a$, which is available at $c$ is now also
available at $d$ (since it is now part of $\mathtt{ctx}$ of thread
2). Thus, the application of interference $ \rf{a}{d} $ becomes redundant.
As a matter of fact, the interference $ \rf{a}{d} $ turns out to be 
infeasible at $d$. This is because {\em extending} the $PO_x$ at $d$ with
the $PO_x$ at $a$ (by taking the {\em meet} of the two orders,
see \textsection\ref{sec:mo-po-lat})
breaks the acyclicity of $PO_x$ at $d$ -- one can visualise this
by adding an edge from $a$ to itself in the Hasse diagram of the
resulting order).  In general, to
address this issue of invalid application of effects at a state, we
introduce the {\em valid extensionality} check
(see \textsection\ref{sec:mo-po-lat}).
Thus, maintaining states this way avoids the need
to perform expensive {\em interference infeasibility} checks. Notably,
such expensive checks are used by other techniques for precision, such as
FruitTree~\cite{KusanoWang-FSE17}.

After two iterations, a fix-point is reached. We can now observe that
at $d$ there is only a single state reachable when $r1 = 1$, which is:
$(PO_x, V_x) = ((\{\mathtt{a}\}, \emptyset), 1), (PO_y, V_y) =
((\{\mathtt{b} \}, \emptyset), 1)$. Thus the property $P$ is shown to be valid
by our analysis.

\subsection{Over-approximating \PO Domain}
Posets are {\em history-preserving} and their use lends precision
to our analysis, however, at the expense of possibly maintaining many
posets. 
We show through a simple example that with further abstraction of forgetting
older sb-ordered stores in the posets (see \hyperlink{c2}{C2}) one can 
obtain succinct posets, thereby resulting 
\begin{wrapfigure}{l}{0.3\linewidth}
\tikzset{every picture/.style={line width=0.75pt}} 
\begin{tikzpicture}[x=1em,y=1em,yscale=-1,xscale=-1]
\tikzstyle{every node}=[font=\footnotesize]

\node (initx) at (0,0) {$ $};

\node (sigE) [left=42pt of initx] {$ $};
\node (poEn1) [circle,fill=black,inner sep=0pt,minimum size=3pt, above right=5pt and 8pt of sigE] {};
\node (Ea) [left=1pt of poEn1, inner sep=0pt] {$a$};
\node (poEn2)[circle,fill=black,inner sep=0pt,minimum size=3pt, below=7pt of poEn1] {};
\node (Ed) [left=1pt of poEn2, inner sep=0pt] {$d$};
\node (poEn3)[circle,fill=black,inner sep=0pt,minimum size=3pt, below=7pt of poEn2] {};
\node (Eb) [left=1pt of poEn3, inner sep=0pt] {$b$};
\draw [color=black ] (poEn1) -- (poEn2);
\draw [color=black ] (poEn2) -- (poEn3);
\node (poE) [draw,fit=(poEn1)(Ea)(poEn2)(Eb)(poEn3)(Ed), color=orange, thin, inner sep=1pt] {};

\node (sigE) [left=8pt of initx] {$$};
\node (poEn1) [circle,fill=black,inner sep=0pt,minimum size=3pt, above right=0pt and 5pt of sigE] {};
\node (Ea) [left=1pt of poEn1, inner sep=0pt] {$a$};
\node (poEn2)[circle,fill=black,inner sep=0pt,minimum size=3pt, right=8pt of poEn1] {};
\node (Ed) [right=1pt of poEn2, inner sep=0pt] {$d$};
\node (poEn3)[circle,fill=black,inner sep=0pt,minimum size=3pt, below left=7pt and 3pt of poEn2] {};
\node (Eb) [left=1pt of poEn3, inner sep=0pt] {$b$};
\draw [color=black ] (poEn1) -- (poEn3);
\draw [color=black ] (poEn2) -- (poEn3);
\node (poE) [draw,fit=(poEn1)(Ea)(poEn2)(Eb)(poEn3)(Ed), color=orange, thin, inner sep=1pt] {};

\node (sigEAbs) [right=20pt of initx] {$ $};
\node (poEn1Abs) [circle,fill=black,inner sep=0pt,minimum size=3pt, above right=-6pt and 5pt of sigEAbs] {};
\node (EdAbs) [left=1pt of poEn1Abs, inner sep=0pt] {$d$};
\node (poEn2Abs)[circle,fill=black,inner sep=0pt,minimum size=3pt, below=7pt of poEn1Abs] {};
\node (EbAbs) [left=1pt of poEn2Abs, inner sep=0pt] {$b$};
\draw [color=black ] (poEn1Abs) -- (poEn2Abs);
\node (poEAbs) [draw,fit=(poEn1Abs)(EdAbs)(poEn2Abs)(EbAbs), color=orange, thin, inner sep=1pt] {};
\node (sigEStateAbs) [fit=(poEAbs)(sigEAbs), inner sep=1pt] {}; 
\end{tikzpicture} 
\caption{Two posets and an abstract poset}
\label{fig:intlv-states}
\end{wrapfigure}
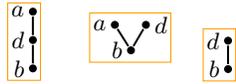
%
in fewer abstract states,
in many scenarios.
%
%
%
%
%
Consider the two example posets (leftmost and center) on variable $x$ 
denoting two distinct states at a location in a
program as shown in
Figure~\ref{fig:intlv-states}.  
%
%
Assume that stores $a$ and $b$ are $\color{brown}{sb}$-ordered and store $d$ is from a
different thread. 
By forgetting the older $\color{brown}{sb}$-ordered store $a$, a smaller abstract $PO_x$
is obtained, which is shown as the rightmost poset in the figure. Notice
that for two distinct states with differing posets at a location, the same abstract poset
is obtained; consequently a single abstract state. 
 This results in a smaller 
abstract state graph.  However, if the value of store at $a$ was read
in a variable that affected an assertion, then
the over-approximated abstract state could result in a loss of
 precision leading to a possible false positive. A detailed
 example program corresponding the illustrated example posets can be found in
 Appendix~\ref{appendix:abs-po}.

\section{Preliminaries} \label{sec:prelim}
\noindent {\bf \RA semantics.}
Given a multithreaded program $P:= \parallel_{i\in \mathtt{Tid}} P_i$,
where $\mathtt{Tid}= \{1,\ldots,n\}$ is the set of thread ids and
$\parallel$ is a parallel composition operator.
Let $ \vars, $ and $ \mathcal{L} $ be the set of shared
variables and set of program locations, respectively.
We use $ (\lab, i) $ to denote the event corresponding to the $ i^{th} $ 
occurrence of program instruction labeled $ \lab $.
Let $ \mathtt{St} $, $ \mathtt{Ld} $ and $ \mathtt{RMW} $ be
the set of all store, load and rmw (read-modify-write) events from $P$, respectively.
%
We denote relations sequenced-before and reads-from 
of  \RA  model \cite{LahavVafeiadis-ICALP15,Batty-POPL11} by
$\seqb{}{}$ and $\rf{}{}$, respectively .  The notation
$ \seqb{a}{b} $ and $ \rf{s}{l} $ represents $ (a,b) \in \seqb{}{} $
and $ (s,l) \in
\rf{}{} $, respectively.
The {\em happens-before} ({\tt hb}) relation for \RA concurrency
is defined as a transitive closure $(\rf{}{} \cup
\seqb{}{})^{+}$.
Let $(\modord{M}{x}, \mo{\modord{M}{x}})$
denote the {\em modification order} ({\tt mo}) over a set of store and rmw events
$ \modord{M}{x} \subseteq \mathtt{St} \cup \mathtt{RMW} $ to a memory location $x$ 
in a program execution.
As defined in \cite{LahavVafeiadis-ICALP15,Batty-POPL11}, 
every valid \RA program execution must have a \texttt{mo} that is consistent with \texttt{hb}.

\noindent {\bf Loset.} The total ordering relation $\mo{\modord{M}{x}}$ is a relation between
every pair of stores $w_1, w_2 \in \modord{M}{x}$ in a program execution
 (alternatively represented as $
\inmo{\modord{M}{x}}{w_1}{w_2} $).
%
We alternatively refer to a modification order as a {\em loset} (linearly
ordered set).
%
Let $ M^{S}$ be the the set of all possible linear orderings over the
set $ S \subseteq \mathtt{St} \cup \mathtt{RMW} $.
Let $ L (S, \preccurlyeq)$ be a function that gives all possible
linearizations of elements in $ S \subseteq \mathtt{St} \cup \mathtt{RMW} $ 
that respect the set of ordering
constraints $ \preccurlyeq$ (note the difference with $\mo{}{}$).  For example $ L(\{a,b\},\emptyset) $
will result in 
$ \{ \{(a,b)\}, \{(b,a)\}
\} $.  Similarly, $ L(\{a,b,c\}, \{(a,b),(a,c)\}) $ will produce: $ \{(a,b),$ $(a,c),(b,c)\} $ and $\{(a,b),(a,c),$
$(c,b)\} $.

\noindent{\bf Interference.}
Following the description of interferences in \textsection\ref{sec:ovw},
we define interference as a relation $ \mathcal{I} \subseteq \mathtt{Tid}
\times \mathtt{Ld} \times (\mathtt{St} \cup \mathtt{RMW})$ such that
$ \mathcal{I}(t)(\ld) \defeq \mathtt{ctx} \cup \mathtt{St} \cup \mathtt{RMW}
 $, 
$\mathtt{ctx}$ is the store
in the program state at some label in $ pre(\ld) $ for thread $t$.
We define $pre(\ld)$ as the set of labels immediately preceding
$\ld$ in $\color{brown}{sb}$ order. 

\section{Concrete Semantics} \label{sec:con-sem}
We consider the set of {\tt mo} losets per
global variable as concrete semantics of a program. Evidently, the set of {\tt mo}
losets is already a sound over-approximation of the set of concrete executions (see
Defn. 5 in \cite{LahavVafeiadis-ICALP15}). Thus, considering the set
of {\tt mo} losets as concrete program semantics does not break the
soundness of our analysis framework  \cite{CousotCousot-2001};
in fact, it serves the purpose of keeping the concrete semantics 
expressible enough 
while maintaining the ease of further abstractions.

\subsection{Modification Orders as Posets} \label{subsec:mo-lattice}

We define the concrete/collecting semantics by the set $ \conla $ such
 that each element $t \in \conla $ is a 
 subset
 of $ M^{S} $ where $ S \subseteq \mathtt{St} \cup \mathtt{RMW} $.
%
%
Let $ t_1 \defeq (S_1, \totordset{S_1}{n}) $ and 
$ t_2 \defeq (S_2, \totordset{S_2}{m}) $ be two elements of $\conla$, where $ \totordset{S}{n} $ denotes a set of losets 
over $ S $ i.e.  $ \totordset{S}{n}  = \{ \mo{1}, \mo{2}, \ldots \}$.
Two elements $t_1, t_2 \in \conla$ are related by an
ordering relation $ \conleq $, 
denoted by  $ t_1 \conleq t_2$. The definition of the ordering relation is
as follows.

\begin{figure}[!t]
	\subfloat[]
	{\begin{minipage}[b]{0.6\textwidth}
			\centering
			\tikzset{every picture/.style={line width=0.75pt}} 
\begin{tikzpicture}[x=1em,y=1em,yscale=-1,xscale=-1]
\node (t1m1n2) [circle,fill=black,inner sep=0pt,minimum size=3pt] {};
\node (t1m1b) [left=1pt of t1m1n2, inner sep=0pt] {$b$};
\node (t1m1n1) [above =9pt of t1m1n2, circle,fill=black,inner sep=0pt,minimum size=3pt] {};
\node (t1m1a) [left=1pt of t1m1n1, inner sep=0pt] {$a$};
\node (t1m1n3) [below=9pt of t1m1n2, circle,fill=black,inner sep=0pt,minimum size=3pt] {};
\node (t1m1c) [left=1pt of t1m1n3, inner sep=0pt] {$c$};
\draw [color=purple] (t1m1n1) -- (t1m1n2);
\draw [color=purple] (t1m1n2) -- (t1m1n3);
\node (mot1m1) [fit=(t1m1n1)(t1m1a)(t1m1n2)(t1m1b)(t1m1n3)(t1m1c)] {};
\draw [decorate,decoration={brace,amplitude=4pt,mirror,raise=5pt},yshift=0pt] (t1m1a.north) --(t1m1c.south) {};

\node (t1m1Name) [below left=1pt and -8pt of t1m1n3] {$m_{11},$};

\node (t1m2n2) [right=10pt of t1m1n2, circle,fill=black,inner sep=0pt,minimum size=3pt] {};
\node (t1m2c) [left=1pt of t1m2n2, inner sep=0pt] {$c$};
\node (t1m2n1) [above =9pt of t1m2n2, circle,fill=black,inner sep=0pt,minimum size=3pt] {};
\node (t1m2a) [left=1pt of t1m2n1, inner sep=0pt] {$a$};
\node (t1m2n3) [below=9pt of t1m2n2, circle,fill=black,inner sep=0pt,minimum size=3pt] {};
\node (t1m2b) [left=1pt of t1m2n3, inner sep=0pt] {$b$};
\draw [color=purple] (t1m2n1) -- (t1m2n2);
\draw [color=purple] (t1m2n2) -- (t1m2n3);
\node (mot1m2) [fit=(t1m2n1)(t1m2a)(t1m2n2)(t1m2c)(t1m2n3)(t1m2b)] {};
\draw [decorate,decoration={brace,amplitude=4pt,raise=5pt},yshift=0pt] (t1m2n1.north) --(t1m2n3.south) {};

\node (t1m2Name) [below left=1pt and -10pt of t1m2n3] {$m_{12}$};

\node (t1Name) [left = 10pt of t1m1n2]  {$t_1 \assign $};
\node (t1) [fit=(t1Name)(mot1m1)(mot1m2)] {};

%
\node (subset) [right =2pt of t1] {$\conleq$};

\node (t2Name) [right=-3pt of subset] {$t_2 \assign $};

\node (t2m1n2) [right=10pt of t2Name, circle,fill=black,inner sep=0pt,minimum size=3pt] {};
\node (t2m1b) [left=1pt of t2m1n2, inner sep=0pt] {$b$};
\node (t2m1n1) [above =9pt of t2m1n2, circle,fill=black,inner sep=0pt,minimum size=3pt] {};
\node (t2m1a) [left=1pt of t2m1n1, inner sep=0pt] {$a$};
\node (t2m1n3) [below=9pt of t2m1n2, circle,fill=black,inner sep=0pt,minimum size=3pt] {};
\node (t2m1c) [left=1pt of t2m1n3, inner sep=0pt] {$c$};
\draw [color=purple] (t2m1n1) -- (t2m1n2);
\draw [color=purple] (t2m1n2) -- (t2m1n3);
\node (mot2m1) [fit=(t2m1n1)(t2m1a)(t2m1n2)(t2m1b)(t2m1n3)(t2m1c)] {};
\draw [decorate,decoration={brace,amplitude=4pt,mirror,raise=5pt},yshift=0pt] (t2m1a.north) --(t2m1c.south) {};

\node (t2m1Name) [below left=1pt and -8pt of t2m1n3] {$m_{21},$};

) [right=-2pt of t2m1n3] {,};

\node (t2m2n2) [right=10pt of t2m1n2, circle,fill=black,inner sep=0pt,minimum size=3pt] {};
\node (t2m2c) [left=1pt of t2m2n2, inner sep=0pt] {$c$};
\node (t2m2n1) [above =9pt of t2m2n2, circle,fill=black,inner sep=0pt,minimum size=3pt] {};
\node (t2m2a) [left=1pt of t2m2n1, inner sep=0pt] {$a$};
\node (t2m2n3) [below=9pt of t2m2n2, circle,fill=black,inner sep=0pt,minimum size=3pt] {};
\node (t2m2b) [left=1pt of t2m2n3, inner sep=0pt] {$b$};
\draw [color=purple] (t2m2n1) -- (t2m2n2);
\draw [color=purple] (t2m2n2) -- (t2m2n3);
\node (mot2m2) [fit=(t2m2n1)(t2m2a)(t2m2n2)(t2m2c)(t2m2n3)(t2m2b)] {};

\node (t2m2Name) [below left=1pt and -11pt of t2m2n3] {$m_{22},$};

\node (t2m3n2) [right=10pt of t2m2n2, circle,fill=black,inner sep=0pt,minimum size=3pt] {};
\node (t2m3a) [left=1pt of t2m3n2, inner sep=0pt] {$a$};
\node (t2m3n1) [above =9pt of t2m3n2, circle,fill=black,inner sep=0pt,minimum size=3pt] {};
\node (t2m3c) [left=1pt of t2m3n1, inner sep=0pt] {$c$};
\node (t2m3n3) [below=9pt of t2m3n2, circle,fill=black,inner sep=0pt,minimum size=3pt] {};
\node (t2m3b) [left=1pt of t2m3n3, inner sep=0pt] {$b$};
\draw [color=purple] (t2m3n1) -- (t2m3n2);
\draw [color=purple] (t2m3n2) -- (t2m3n3);
\node (mot2m3) [fit=(t2m3n1)(t2m3c)(t2m3n2)(t2m3a)(t2m3n3)(t2m3b)] {};
\draw [decorate,decoration={brace,amplitude=4pt,raise=5pt},yshift=0pt] (t2m3n1.north) --(t2m3n3.south) {};

\node (t2m3Name) [below left=1pt and -11pt of t2m3n3] {$m_{23}$};
\node (t2) [fit=(t2Name)(mot2m1)(mot2m2)(mot2m3)] {};

\end{tikzpicture}
			\label{fig:mo-ordering1}
			\tikzset{every picture/.style={line width=0.75pt}} 
\begin{tikzpicture}[x=1em,y=1em,yscale=-1,xscale=-1]

\node (t3m1n1) [circle,fill=black,inner sep=0pt,minimum size=3pt] {};
\node (t3m1a) [left=1pt of t3m1n1, inner sep=0pt] {$a$};
\node (t3m1n2) [below =9pt of t3m1n1, circle,fill=black,inner sep=0pt,minimum size=3pt] {};
\node (t3m1b) [left=1pt of t3m1n2, inner sep=0pt] {$b$};
\draw [color=purple] (t3m1n1) -- (t3m1n2);
\node (mot3m1) [fit=(t3m1n1)(t3m1a)(t3m1n2)(t3m1b)] {};
\draw [decorate,decoration={brace,amplitude=2pt,mirror,raise=3pt},yshift=0pt] (t3m1a.north) --(t3m1b.south) {};
\draw [decorate,decoration={brace,amplitude=2pt,raise=3pt},yshift=0pt] (t3m1n1.north) --(t3m1n2.south) {};
\node (t3m1Name) [below left=1pt and -8pt of t3m1n2] {$m_{31}$};

\node (t3Name) [below left =1pt and 10pt of t3m1n1]  {$t_3 \assign $};
\node (t3) [fit=(t3Name)(mot3m1)] {};

\node (subset) [right =2pt of t3] {\normalsize $\conleq$};

\node (t4Name) [right =2pt of subset] {$t_4 \assign $};

\node (t4m1n1) [right =9pt of t4Name, circle,fill=black,inner sep=0pt,minimum size=3pt] {};
\node (t4m1a) [left=1pt of t4m1n1, inner sep=0pt] {$\{a$};
\node (t4m1cb) [right=-3pt of t4m1n1] {$\}$};
\node (mot4m1) [fit=(t4m1n1)(t4m1a)] {};

\node (t2m3Name) [below left=1pt and -9pt of t4m1n1] {$m_{41}$};
\node (t4) [fit=(t4Name)(mot4m1)] {};

\end{tikzpicture}
			\label{fig:mo-ordering2}
	\end{minipage}}
	\subfloat[]
	{\begin{minipage}[b]{0.3\textwidth}
			\tikzset{every picture/.style={line width=0.75pt}} 
\begin{tikzpicture}[x=1em,y=1em,yscale=-1,xscale=-1]

\node (poJoinN1) [circle,fill=black,inner sep=0pt,minimum size=3pt] {};
\node (poJoinC) [left=1pt of poJoinN1, inner sep=0pt] {$c$};
\node (poJoin) [draw,fit=(poJoinN1)(poJoinC), color=orange, thin, inner sep=1pt] {};
\node (poJoinName) [left=0pt of poJoin] {$p_{\join} = $};

\node (join) [below=0pt of poJoin] { $\join$};

\node (p1Name) [below left=15pt and 17pt of join] {$p_1 = $};
\node (p1N1) [above right=-2pt and 8pt of p1Name, circle,fill=black,inner sep=0pt,minimum size=3pt] {};
\node (p1a) [left=1pt of p1N1, inner sep=0pt] {$a$};
\node (p1N2)[ below=7pt of p1N1, circle,fill=black,inner sep=0pt,minimum size=3pt] {};
\node (p1c) [left=1pt of p1N2, inner sep=0pt] {$c$};
\draw [color=black ] (p1N1) -- (p1N2);
\node (p1po) [draw,fit=(p1N1)(p1a)(p1N2)(p1c), color=orange, thin, inner sep=1pt] {};

\node (p2Name) [below right=15pt and -16pt of join] {$p_2 = $};
\node (p2N1) [above right=-2pt and 8pt of p2Name, circle,fill=black,inner sep=0pt,minimum size=3pt] {};
\node (p2b) [left=1pt of p2N1, inner sep=0pt] {$b$};
\node (p2N2)[below=7pt of p2N1, circle,fill=black,inner sep=0pt,minimum size=3pt] {};
\node (p2c) [left=1pt of p2N2, inner sep=0pt] {$c$};
\draw [color=black ] (p2N1) -- (p2N2);
\node (p2po) [draw,fit=(p2N1)(p2b)(p2N2)(p2c), color=orange, thin, inner sep=1pt] {};

\draw (join.south west) -- (p1po.north);
\draw (join.south east) -- (p2po.north);

\node (meet) [below=42pt of join] { $\meet$};
\draw (meet.north west) -- (p1po.south);
\draw (meet.north east) -- (p2po.south);

\node (poMeetN1) [below left=0pt and -2pt of meet, circle,fill=black,inner sep=0pt,minimum size=3pt] {};
\node (poMeeta) [left=1pt of poMeetN1, inner sep=0pt] {$a$};
\node (poMeetN2) [right=8pt of poMeetN1, circle,fill=black,inner sep=0pt,minimum size=3pt] {};
\node (poMeetb) [right=1pt of poMeetN2, inner sep=0pt] {$b$};
\node (poMeetN3) [below left=7pt and 3pt of poMeetN2, circle,fill=black,inner sep=0pt,minimum size=3pt] {};
\node (poMeetc) [left=1pt of poMeetN3, inner sep=0pt] {$c$};
\draw [color=black ] (poMeetN1) -- (poMeetN3);
\draw [color=black ] (poMeetN2) -- (poMeetN3);
\node (poMeet) [draw,fit=(poMeetN1)(poMeeta)(poMeetN2)(poMeetb)(poMeetN3)(poMeetc), color=orange, thin, inner sep=1pt] {};
\node (poMeetName) [left =0pt of poMeet] {$p_{\meet} = $};

\end{tikzpicture}
			\label{fig:po-oredring}
	\end{minipage}}
	\caption{Orderings over $ \conla, \mathcal{P} $}
\end{figure}
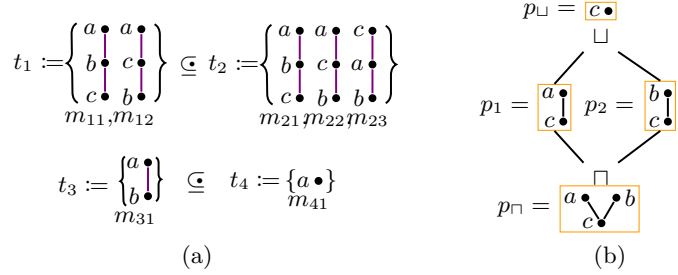

\begin{dfn} 
$ t_1 \conleq t_2 \iff (S_1 \supseteq
S_2 \andspace \forall \mo{i} \in \totordset{S_1}{n}  \exists \mo{j} \in \totordset{S_2}{m}\
. \forall a,b \in S_2\ a \mo{i} b \implies a \mo{j} b) $.
\end{dfn}
We extend the set $ \conla $ with a special element $ \bot_{\conla} $ such that 
$ \forall t \in \conla\ .\ \bot_{\conla} \conleq t $. 
Each element in 
$ \conla $ is a set of {\tt mo} losets that represents a set of (possibly
partial) executions. 
For instance, $ t_1 $  in Figure~\ref{fig:mo-ordering1}
is an over-approximation of all the executions whose {\tt mo}s 
satisfy either $m_{11}$ or $m_{12}$.
Note  $ t_1 \conleq t_2 $, 
which means that
the set of executions corresponding 
to $ t_2 $ is  larger than the set of executions corresponding to $ t_1 $.
We infer that
$ t_1 $ gives us more precise information on execution possibilities
than $ t_2 $ for the same program.
Similarly, in Figure~\ref{fig:mo-ordering2} element $t_3$ is ordered below
$t_4$. The set of executions having $m_{41}$ 
as a part of their {\tt mo} is larger than set of executions having
$m_{31}$ as part of their {\tt mo}.

The element $ \bot_{\conla} $ represents a set
in which all modification orders are inconsistent, and hence
represents an invalid execution.  Likewise, we introduce element $ \top_\conla =
(\emptyset, \emptyset) $  in the $ \conla $ representing an empty set of
constraints, which is equivalent to the set of all valid
executions. By definition, $ \top_\conla $ is ordered above all the
elements $ \conla $ in the $ \conleq $. We establish that $\conla$ is a
poset under the relation $\conleq$.

\begin{lemma}\label{lem:t-complete-l}
	$ (\conla, \conleq) $, is a poset.\footnote{Proofs of all lemmas and theorems in this
		article are available in
		the Appendix~\ref{appendix:lo-po} \label{fn:fnt1}}
\end{lemma}

\section{Abstract Semantics} \label{sec:abs-sem}
We present a 
two-layered abstraction to arrive at final
abstract \RA program semantics. In particular, (i) the
set of {\tt mo} losets of a program is abstracted in to \PO domains, 
and (ii) the 
\PO domains are further over-approximated, where for each
variable all stores older than the latest store under {\tt sb} ordering in its
poset are forgotten.
Further, we demonstrate that abstract semantics produced in step (i) from above
 forms a {\em complete lattice} and establish  
a  {\em  Galois connection} between the concrete and abstract domains.

\subsection{{\tt mo} Posets as Lattices}
\label{sec:mo-po-lat}

%
In this section we define a lattice over
$ \mathcal{P} $ which is the set of all
partial orders. We use the terms {\tt mo} poset and \PO domain
interchangeably for this lattice.

We combine two or more {\tt mo} losets and
respresent them as a collection of
partial orders.
For instance, consider {\tt mo} losets $ p_1$ and $p_2$ (shown in
Figure \ref{fig:po-oredring}) in $\mathcal{P}$. These can be combined in the
following two ways: (i) the orderings in $p_1$ and $p_2$ are both
present in the combination (the binary operator is denoted by $\meet$
and the resulting element is denoted by $p_{\meet}$), or (ii) common
orderings in $p_1$ and $p_2$ on the common elements are present in the
combination (the binary operator is denoted by $\join$ and the resulting element
is denoted by $p_{\join}$).
After the application of step (i), we note that the pairs $(a,b)$ or
$(b,a)$ are not in the relation $ p_1 \meet p_2 $. Similarly, after
the application of step (ii), we note that all those executions that contain $c$
are included in $p_{\join}$. Also, note that $p_{\meet},
p_{\join} \in \mathcal{P}$. 
Going forward we define the following operations over the elements in a set of partial orders: 
\begin{description}
	\item[Less ($ p_1 \leqlattice p_2 $):] 
	An ordering relation among two partial orders
	$ p_1 = (\modord{M}{x},\pomo{1})) $ and $ p_2 = (\modord{N}{x},\pomo{2}))$, $ p_1, p_2 \neq \bot $
	is defined as following:
	$ p_1 \leqlattice p_2 \iff
	\modord{M}{x} \supseteq \modord{N}{x} \andspace
	\inpomo{2}{a}{b} \implies
	\inpomo{1}{a}{b}) $ and $ \forall p \in \mathcal{P}, \bot \leqlattice p $
	
	\item[Is Consistent ($ p_1 \iscons p_2$):] Two partial orders
	are consistent with each other if they do not contain any {\em
	conflicting} pair and $ \bot $ is not consistent with any
	element.  Formally, $ \bot \iscons p_2 \defeq false $, $
	p_1 \iscons \bot \defeq false $ and $ \forall p_1,
	p_2 \neq \bot $, $ p_1 \iscons p_2 \defeq \forall
	a,b \in \modord{M}{x} \cup \modord{N}{x} \ .\ a \neq b, \
	(a,b) \in\ \pomo{1} \implies (b,a) \notin \ \pomo{2} $. We
	denote inconsistent partial orders using the notation $
	p_1 \isncons\ p_2 $.
	
	
	\item[Is Valid Extension($p \isext \st$):] A store event $ \st $ is a 
	valid extension of the partial order $p = (\modord{M}{x},\pomo{})$
        {\em iff} there is no 
	instruction ordered after $ \st $ in the ordering relation $\pomo{}$.
	Formally, $ p \isext \st \defeq \forall a\in \modord{M}{x},
        (st,a) \notin\ \pomo{} $.
        A invalid extension of a partial order $p$ by a store $st$ is denoted by
        $p \isnext \st$).
	
	
	\item[Append ($ p \append \st $):] Appends the store operation $ \st $ at 
	the end of modification order $ p = (\modord{M}{x},\pomo{}) $
        if $ \st $ is a valid extension of $ p $ i.e. 
	$ p \append \st \defeq $ if $ p \isext \st $ then $ (\modord{M}{x} \cup \{\st\}, 
	\pomo{} \cup\ \{(a, st)\mid a \in\ \modord{M}{x} \}) $ else $ \bot $.
	
    \item[Meet ($ p_1 \meet p_2$):] The meet of two partial orders
      $p_1 = (\modord{M}{x},\pomo{1}))$ and
        $p_2= (\modord{N}{x},\pomo{2}))$ is defined as:
	$ p_1 \meet p_2 \defeq $ if $ p_1 \iscons p_2 $ then $ (\modord{M}{x} \cup 
	\modord{N}{x}, \pomo{1} \cup \pomo{2}) $ else $ \bot $.
	
	\item[Join ($ p_1 \join p_2$):] The join of two partial order
        $ p_1 = (\modord{M}{x},\pomo{1})) $ and $ p_2 = (\modord{N}{x},\pomo{2})$, $ p_1, p_2 \neq \bot $  
	is defined as the intersection of common ordered pairs  in the partial orders, i.e, 
	$ p_1 \join p_2 \defeq (\modord{M}{x} \cap \modord{N}{x}, \pomo{1} 
	\cap \pomo{2})$. 
	We define $ \bot \join p_2 \defeq p_2 $ and $ p_1 \join \bot \defeq p_1 $.
	
	\item[Widening ($ p_1 \widen p_2 $):] The widening operator over two partial orders	
	 $ p_1 = (\modord{M}{x},$ $ \pomo{1}) $ and $ p_2 = (\modord{N}{x},\pomo{2})$, 
	 $ p_1, p_2 \neq \bot $ is defined as
	 $ p_1 \widen p_2 \defeq (\modord{Q}{x}, \pomo{}) $, where 
	 $ \modord{Q}{x} = \{ a \mid a = (\lab,i) \in \modord{M}{x} \cap \modord{N}{x} \andspace 
	 	\nexists b=(\lab,j) \in \modord{M}{x} \cap \modord{N}{x} \ .\ j < i \} $ 
	 and $ \pomo{} = \{ (a,b) \mid (a,b) \in \pomo{1} \cap \pomo{2} \andspace \ a,b \in \modord{Q}{x} \} $.
	 We define $ \bot \widen p_2 \defeq p_2 $ and $ p_1 \widen \bot \defeq p_1 $.
          
\end{description}

\begin{lemma} \label{lem:lub-glb-po}
The operators $ \join $ and $\meet$ define the lub and glb of any two
elements of $\mathcal{P}$, respectively.\footnoteref{fn:fnt1}
\end{lemma}


\begin{lemma}\label{lem:p-complete-l}
$ (\mathcal{P}, \leqlattice, \join, \meet, \bot, \top) $ is a complete
  lattice, where $ \mathcal{P} $ is set of all possible partial orders
  over elements of set $ \mathtt{St} \cup \mathtt{RMW} $, $ \top $ is defined as
  empty poset, and $ \bot $ is a special element
  that is ordered below all the elements of $ \mathcal{P} $ in $
  \leqlattice $.\footnoteref{fn:fnt1}
\end{lemma}
The proof of Lemma~\ref{lem:p-complete-l} follows from Lemma \ref{lem:lub-glb-po}, 
the definition of $\join$ and $\meet$ operations 
of $\mathcal{P}$, and standard properties of operators.

\begin{lemma} \label{lem:widening}
	The binary operation $ \widen $ defines a widening operator over the elements of the lattice 
	$ (\mathcal{P}, \leqlattice, \join, \meet, \bot, \top)$.\footnoteref{fn:fnt1}
\end{lemma}

We explain the widening operator $ \widen $ using an example.
Recall that each element of lattice $ \mathcal{P} $ is a partial order over 
program events. 
Let $ p = (\modord{Q}{x},\pomo{} ) = p_1 \widen p_2 $, then the set of events in $ p $ 
 maintains the earliest occurrence of common events in
$ \modord{M}{x}$ and $ \modord{N}{x} $ corresponding to $ p_1 $ and $
p_2 $, respectively. Consider the events $ e_2 = (\lab, 2) $ and $ e_3
= (\lab, 3) $, which are generated by the same program instruction
labeled $ \lab $. If both $ p_1 $ and $ p_2 $ contain the ordering
 $e_2$ and $e_3$, then the result of widening will
contain the earliest occurrence of an event from $\lab$, i.e., $ e_2 $
so long as $ e_1 = (\lab, 1) \notin \modord{M}{x} \cap \modord{N}{x}
$.  The set of orderings $ \pomo{} $ is defined over
the elements of $ \modord{Q}{x} $. Hence no ordering involving $ e_3
$ in this example will be in $ \pomo{} $.

Given a monotone function $ f:\mathcal{P} \rightarrow \mathcal{P} $, 
consider the chain $ f_{\widen}^0, f_{\widen}^1, f_{\widen}^2 \dots $
with $ f_{\widen}^0 = \bot $ and 
$ f_{\widen}^i = f_{\widen}^{i-1} \widen f(f_{\widen}^{i-1}) $ for some $ i>0 $. 
An essential requirement on  $ \widen $ for it to be a widening operator 
 is that the above chain must stabilize, i.e.,  
$ f(f_{\widen}^{n}) \leqlattice f_{\widen}^{n} $ for some $ n>0 $. It means 
that the function $ f $ is {\em reductive} at $ f_{\widen}^{n} $.
We show in the proof of Lemma~\ref{lem:widening} that 
our defined operator $ \widen $ is indeed a widening operator. Using Tarski's 
fixpoint theorem, it follows that $ \mathtt{lfp}(f) \leqlattice f_{\widen}^{n} $, 
where $ \mathtt{lfp}(f) $ is the least fixed point of $ f $. 
As a result, $ f_{\widen}^{n} $ is a sound over-approximation of $ f $, which
guarantees termination of analysis with infinite lattices having infinite ascending chains.

%
%

\begin{dfn}
The abstraction function
$ \alpha: \mathcal{T} \rightarrow \mathcal{P} $ is defined as
$ \alpha(\bot_{\conla}) \defeq \bot $ and $ \forall t \neq \bot_{\conla} $,
$ \alpha(t) \defeq (\modord{M}{x}, \pomo{}) $
for some $ t =
(S, \totordset{}{n}) $ given $ \modord{M}{x} = S $, and
$ \pomo{}
= \intersectionof{}{} \mo{i} $.
\end{dfn}

\begin{dfn}
The concretization function
$ \gamma: \mathcal{P} \rightarrow \mathcal{T} $ is defined as
$ \gamma(\bot) \defeq \bot_{\conla} $ and $ \forall p \neq \bot $,
$ \gamma(p) \defeq (S, \totordset{}{n}) $ for some $ p
=(\modord{M}{x}, \pomo{}) $ given $ S = \modord{M}{x} $ and
$ \totordset{}{n} $ is set of all possible linearizations of
$ \pomo{} $ i.e. $ \totordset{}{n} = L(S, \pomo{})
$.
\end{dfn}
 Having defined the abstraction and concretization operators, we can now establish
 the Galois connection between the poset $\mathcal{T}$ and the lattice $\mathcal{P}$.
\begin{theorem} \label{thm:to-po-galois}
$	(\mathcal{T}, \conleq) \galois{\alpha}{\gamma}
(\mathcal{P}, \leqlattice, \join, \meet, \bot, \top) $. \footnoteref{fn:fnt1}
\end{theorem}

We lift the result from Theorem~\ref{thm:to-po-galois} to the product lattices of all the program variables. Theorem~\ref{thm:pointwise-galois} articulates that the Galois connection between concrete and abstract product lattices
is preserved.
\begin{theorem} \label{thm:pointwise-galois}
The correspondence between 
	$ \prod_{x \in \vars} \mathcal{P}_x $ and 
	$ \prod_{x \in \vars} \mathcal{T}_x $ is a Galois connection. \footnoteref{fn:fnt1}
\end{theorem}

It is worthwhile to note that lattice $\mathcal{P}$ is not tied to the
\RA semantics. As such, the \PO domain is not specific to any memory
model. We present a discussion in \textsection\ref{subsec:generic-mm},
on the applicability of \PO domain beyond \RA semantics.
Below, we give a description of transfer functions 
for the operations in \RA programs. 


\subsection{Abstract Semantics of \RA programs} \label{subsec:sem-ra}
The values of shared variables in the program can be abstracted to any known 
numeric abstract domain such as interval, octagon, or polyhedra. Let 
$ \abs{\mathbb{V}} $ represents the set of values in the chosen abstract domain.
Let $ \mathcal{M}: \vars \rightarrow \abs{\mathbb{V}} $ define the memory state 
of a program. 
%
Let 
$ \mathbb{M}: \vars \rightarrow \mathcal{P} $ represent a
map from shared variables to corresponding elements in the abstract {\tt mo} poset lattice $ \mathcal{P} $.
We abuse notations $ \append, \join, \meet, \widen, \iscons$, and $ \isext $
to represent the corresponding pointwise-lifted operators for $ \mathbb{M} $. 
For instance, the pointwise lifting of $ \append $ appends the stores
of variable $ v $ only to its
modification order ({\em i.e.},
$ \mathbb{M}(v) $); the modification orders $\mathbb{M}(v')$ for 
variables $ v' \neq v $ remain unchanged.
The pointwise lifting for other operators is  straighforward.
From Theorem~\ref{thm:pointwise-galois}, it follows that $\mathbb{M}$ 
	along with the pointwise lifted operators constitute the sought abstract domain.

\begin{figure}[t]
	\centering
	\proofrule{store} 
	{\begin{trgather} (pre(\lab),mo,m) \in \mathcal{S} \qquad m'=m[x \rightarrow v] \\
	 mo'= mo[x \rightarrow mo(x) \append \lab] \end{trgather}}
	{\mathcal{S} \xrightarrow{\lab: \store{x}{v}} \mathcal{S} \squplus (\lab, mo',m')}	
\\ \ \\ \ \\
	\proofrule{rmw}
	{\begin{trgather} (pre(\lab),mo,m) \in \mathcal{S} \\
	(pre(\lab),mo,m) \xrightarrow{\lab: \load{x}} (\lab, mo'',m'') \\ m''(x)=v  \\
	(pre(\lab),mo'',m'') \xrightarrow{\lab: \store{x}{v'}} (\lab, mo',m') \end{trgather}}
	{\mathcal{S} \xrightarrow{\lab: \rmwinst{x}{v}{v'}} \mathcal{S} \squplus (\lab, mo',m')}
\\ \ \\ \ \\	
	\hspace{2pt}	
	\proofrule{load}
	{\begin{trgather} (pre(\lab),mo_l,m_l) \in \mathcal{S} \qquad \st \in \mathcal{I}(t)(\lab) \\
			(\st, mo_s, m_s) \in \mathcal{S} \\ 
			(pre(\lab), mo', m'') = \mathtt{AI}((pre(\lab),mo_l,m_l),(\st, mo_s, m_s)) \\
			m' = m''[x \rightarrow m_s(x)] \end{trgather}}
	{\mathcal{S} \xrightarrow{\lab: \load{x}} \mathcal{S} \squplus (\lab, mo',m')}	

	\caption{Transfer functions for \RA programs. 
        \texttt{AI}($(\lab_1, mo_1, mo_2), (\lab_2,
        mo_2,m_2)$) $\defeq$ $(\lab_1, (mo_1 \append \lab_2) \meet
        mo_2, m_1 \join m_2)$; \texttt{AI} applies the interference
        from $(\lab_2, mo_2,m_2)$ to the memory and {\tt mo} poset 
        state of $(\lab_1, mo_1, mo_2)$.
        } \label{fig:con-sem}
\end{figure}
Let $ \Sigma \subseteq \loc \times (\mathbb{M} \times \mathcal{M}) $
represents the set of all reachable program states. 
%
The transfer functions for operations $ \ld, \st$ and $ \rmw $ are
defined in Figure~\ref{fig:con-sem}. 
We provide additional rules (for lock and unlock)
and auxillary functions, which are supported by our technique, in
Appendix~\ref{appendix:abs-sem}.
Since we assume the SSA
representation of programs, arithmetic operations only modify the
thread local variables. As a result, $\mathbb{M}$ remains
unchanged. The effects of arithmetic operations on shared variables is
captured via numeric abstract domains. Thus, the transfer
functions for such operations are excluded from our presentation.
The semantic definitions in Figure~\ref{fig:con-sem} are parameterized
in terms of the set of currently explored reachable program states,
$ \mathcal{S} \subseteq \Sigma $, at a some point during the analysis.
\newline
%
%
%
\indent Consider the \textsc{load} rule which, defines the semantics of a load
operation.  A load of a shared variable $ x $ at $\lab$ is
performed at program state(s) $ \mathcal{S} $ using the following
steps.  Let $\st$ be an interfering instruction for $\lab$. Each
explored program state $(\st, mo_s, m_s)$ at instruction label $\st$ is
considered as an interference and analyzed with the set of program states at 
label $pre(\lab)$ using the function \texttt{AI} (defined in the caption of
Figure~\ref{fig:con-sem}).
%
%
When the interference from program state $(\lab_2, mo_2,m_2)$ is successfully applied 
to the program state $(\lab_1, mo_1, mo_2)$ by function \texttt{AI} (the load at $\lab_1$ 
reads from the store at $\lab_2$), then as a result $\lab_2$ is appended
in the partial order at $ \lab_1 $, \ie $mo_1$.
%
For all other events prior to $\lab_1$ and $\lab_2$ , the precise
ordering information among them is computed by taking the {\em meet}
of $mo_1$ and $mo_2$, \ie $mo_1 \meet mo_2$ (because the ordering of
such events must be consistent with both $mo_1$ and $mo_2$).

%
%
In the state at $\lab_1$, the value of variables other than interfering variable $x$ can come from either $m_1$ or $m_2$. 
The function \texttt{AI} {\em joins} the maps $m_1$ and $m_2$ to obtain 
all feasible values for such variables. 
To compute $\join$ on memory values, one can choose abstract
domains such as intervals or octagons.
Let \texttt{AI} return $(pre(\lab),mo',m'')$ when the interference 
is applied from $(\st, mo_s, m_s)$ to $(pre(\lab),mo_l,m_l)$.
The value of variable $x$ read by the load operation $\lab$ in the 
program state $(pre(\lab),mo',m'')$ will be the same as the value 
of variable $x$ in the interfering program state $m_s(x)$.
Thus, we substitute $m''(x)$ with $m_s(x)$ to construct the reachable 
program state $(\lab, mo', m')$.
%
\newline
\indent Finally, the resulting  state at $ \lab $ is combined with the currently
existing states by the $ \squplus $ operator. The operator
$ \squplus $ performs instruction-wise join of  states, i.e., 
it joins the memory state of two states if their instruction labels 
and {\tt mo} posets are the same. It also joins the {\tt mo} poses 
if the instruction label and the memory states are the same, otherwise, it 
leaves the two states as is.
Formally, the operation $ \squplus $ replaces any two program states 
(say $ (\lab_1, mo_1, m_1) $ and 
$ (\lab_2, mo_2, m_2) $), with a single program state 
$ (\lab_1, mo_1, m) $, where $ m = m_1 \join m_2 $, if $ mo_1 = mo_2
\andspace \lab_1 = \lab_2 $, and with $ (\lab_1, mo, m_1) $, where 
$ mo = mo_1 \join mo_2 $, if $ m_1=m_2 \andspace \lab_1 = \lab_2 $.

Transfer functions for \textsc{rmw} and \textsc{store} can be interpreted
in a similar way. Readers may note that, in general, two successful
$\mathtt{RMW}$ operations will never read from the same store as
is assumed in our rule. However,
our definition is sound (and simple to understand); we provide a more precise 
definition in \textsection~\ref{subsec:updated-rmw}, which is also implemented in
\tool.

\subsection{Abstracting the Abstraction: Approximating {\tt mo} Posets} \label{subsec:abs-po-mo}
We leverage the ordering rules of the \RA memory model to further
abstract the modification orders. 
Let $p \defeq (\modord{Q}{x}, \pomo{}), p_1 \defeq (\modord{M}{x}, \pomo{1}), 
p_2 \defeq (\modord{N}{x}, \pomo{2}), p_a \defeq (\modord{A}{x}, \pomo{a})$ 
be some elements in $ \mathcal{P} $.
We shall use these definitions whenever $ p, p_1, p_2 $ and $ p_a $ appear 
in definitions and predicates below.

%
\noindent Our abstraction function $ \abs{\alpha}: \mathcal{P} \rightarrow \mathcal{P} $ can be 
defined as follows: $ \abs{\alpha}(\bot) = \bot $ and $ \forall p \neq \bot $,
$ \abs{\alpha}(p) \defeq (\modord{A}{x}, \pomo{a}) $, where 
$ \modord{A}{x} = \modord{Q}{x} \setminus \{a \mid \exists b \in \modord{Q}{x}\ .\ 
\seqb{a}{b} \andspace a \neq b \} $ and 
$\pomo{a} = \pomo{} \setminus \{(a,b) \mid (a,b) \in  
\pomo{} \andspace (a \notin \modord{A}{x} \orspace b \notin 
\modord{A}{x}) \}$. 
\newline
\textbf{Soundness of $ \abs{\alpha} $ Abstraction:} Let relation $ 
\beta \in \wp(\mathcal{P} \times \mathcal{P}) $, where $ \wp $ denotes power set, be defined as  $ \exists p_1, 
p_2 \in \mathcal{P}, (p_1, p_2) \in \beta \iff p_1 = \bot \orspace (\modord{N}{x} \subseteq 
\modord{M}{x} \setminus \{a \mid \exists b \in \modord{M}{x} \ .\ \seqb{a}{b} 
\andspace a \neq b \}\ \andspace \pomo{2} \subseteq 
\pomo{1}) $.
Through Lemma \ref{lem:beta-subset} we establish that our definition of $ \beta $ 
indeed provides a soundness relation.

\begin{lemma} \label{lem:beta-subset}
	$ (p_1, p_2) \in \beta \implies p_1 \leqlattice p_2 $. \footnoteref{fn:fnt1}
\end{lemma}

\begin{lemma} \label{lem:greater-in-beta}
	Abstract soundness assumption holds under $ \beta $, i.e., $ \forall p, 
	p_1, p_2 \in \mathcal{P} . $ $(p, p_1) \in \beta \andspace p_1 
	\leqlattice p_2 \implies (p, p_2) \in \beta $. \footnoteref{fn:fnt1}
\end{lemma}

In other words, Lemma \ref{lem:greater-in-beta} allows us to conclude that if 
$ p_1 $ is a sound over-approximation of $ p $, then every element ordered above $ p_1 $
in lattice $ \mathcal{P} $ is also a sound over-approximation of $ p $ under $ \beta $. 
We shall use Lemmas \ref{lem:beta-subset}-\ref{lem:greater-in-beta} to establish the soundness 
of $ \abs{\alpha} $ in the theorem below.

\begin{theorem} \label{thm:abspo-sound}
	Abstraction relation $ \abs{\alpha} $ is minimal sound abstraction under 
	soundness relation $ \beta $, i.e., $ (p_1, p_2) \in \beta \iff 
	\abs{\alpha}(p_1) \leqlattice p_2 $. \footnoteref{fn:fnt1}
\end{theorem}
The proof of Theorem~\ref{thm:abspo-sound} is obtained by a straightforward
application of the definitions of $ \abs{\alpha} $, $ \beta $ and 
Lemma~\ref{lem:greater-in-beta}. 

We redefine some of the operations described in \textsection\ref{sec:mo-po-lat}
in order to assist with the computation of transfer functions under the 
$ \abs{\alpha} $ abstraction:
\begin{description}
	\item[Is Consistent ($ p_1 \iscons p_2 $):]  
	$ \bot \iscons p_2 \defeq false $, $
	p_1 \iscons \bot \defeq false $ and $ \forall p_1,
	p_2 \neq \bot $
	$ p_1 \iscons p_2 \defeq \forall a,b\ ((a,b) \in\ \pomo{1} 
	\implies \forall \seqb{b}{c} \ .\  (c,a) \notin\ \pomo{2}) 
	\andspace ((a,b) \in\ \pomo{2} \implies \forall \seqb{b}{c}
	\ .\  (c,a) \notin\ \pomo{1}) $. Note that $\seqb{}{}$
        is reflexive.
        As before, we use the 
	notation $ p_1 \isncons p_2 $ when $p_1$ and $p_2$ are inconsistent.
	
	\item[Is Valid Extension($p \isext \st$):] $ p \isext \st \defeq 
	\forall a (st,a) \notin\ \pomo{} \andspace \nexists b \in 
	\modord{Q}{x}\ .\ \seqb{st}{b} $. We use the notation $p \isnext \st$ 
	to indicate that $\st$ is not a valid extension of $p$.
	
	\item[Append ($ p \append \st $):] If $ \st $ is a valid extension of 
	$ p $, then append the store operation $ \st $ at the end of partial 
	order $ p $ and delete the older instructions, if any, i.e. 
	$ p \append \st \defeq $ if $ p \isext \st $ then $ (\modord{Q}{x} 
	\cup \st \setminus \{a \mid \seqb{a}{\st} \}, 
	\pomo{} \cup\ \{(a, st) \mid a \in\ \modord{Q}{x} \}\ 
	\setminus\ \{(a,b) \mid (\seqb{a}{\st} \andspace a \neq \st) \orspace 
	(\seqb{b}{\st} \andspace b \neq \st) \}) $ else $ \bot $.
\end{description}

\noindent\textbf{Over-Approximating the Semantics of \RA programs.} 
%

We use the modified definitions of $ \append, \iscons, \isncons, \isext $ 
and  $ \isnext $ operators to perform analysis under $ \abs{\alpha} $ 
abstraction. The semantics of $ \st, \ld $ and $ \rmw $ operations 
and the set of all program states $ \Sigma $ remain the same as  under 
$ \abs{\alpha} $, as defined in \textsection\ref{subsec:sem-ra}.


\subsection{Posets as a Generic Abstraction} \label{subsec:generic-mm}
In this section, we discuss the possibility of using the lattice 
$ (\mathcal{P}, \leqlattice, \join, \meet, \bot, \top) $ as a generic 
abstraction, and using it for reasoning memory models other than \RA.
As a first step, we  reinvestigate how we define
the collecting semantics for programs under non-\RA memory models.
%
The \texttt{mo} losets
may not be best suited collecting semantics
to reason over programs under 
other memory models.


Consider, for instance, the TSO model. 
The collecting semantics for TSO  model require an
ordering over all the
events of shared variables in the program, 
except
among the store-load pairs of different
variables from the same thread. Thus,  using 
losets as concrete semantics over loads and stores of all the shared variables
in which
the  store-load pair of different variables in a thread can appear
in any order
will suffice. This allows us to capture
\texttt{rfe} (reads-from-external,
\texttt{rfe}=$\setRF \setminus \texttt{po}$) in the loset.
Similarly, considering the PSO model 
 the concrete semantics containing
 one loset per variable containing all the load and store events
of that variable will suffice.


Note that once the collecting semantics
is suitably fixed, then formal objects such as
 $(\mathcal{T}, \conleq)$,  
$(\mathcal{P}, \leqlattice, \join, \meet, \bot, \top)$, and 
functions $\alpha$ and $\gamma$ can be used in the 
analysis without requiring any change.
However, designing $\abs{\alpha}$ for other memory models may require
careful analysis, and is left as future work.

\section{Thread-modular Abstract Interpretation} \label{sec:thr-mod}
\subsection{Analysis Algorithms}
We present Algorithm~\ref{alg:thread-modular} in which 
 procedure \texttt{ThreadModularAnalysis} analyzes the entire program by 
considering one thread at a time.
The analysis begins with the initialization of the set of explored
program states (line 2). For each thread $t \in \mathtt{Tid}$,
relation $ \mathcal{I}(t)$ is computed (line 3) according to the definition
in \textsection\ref{sec:prelim}.
Each thread is analyzed under all possible interferences in $ \mathcal{I} $ until
a fixed point is reached (lines 4-8).
The function $\mathtt{SeqAI(t,\mathcal{S},\mathcal{I}(t))}$ is a
standard work-list based sequential abstract interpretation over a
single thread \cite{Nielson-2010}. Our work adapts this analysis by
replacing the transfer functions with the ones given in
\textsection\ref{subsec:sem-ra}.  The function returns a set of
 states for all the locations in the thread $t$.  The
  operator $ \squplus $ performs instruction-wise join (explained in
  \textsection\ref{subsec:sem-ra}) of environments in the existing
  ($\sigma$, line 5) and the newly computed program states
  ($\mathtt{SeqAI(t,\mathcal{S},\mathcal{I}(t))}$).
The details of \RA memory model,
interferences, abstractions and semantics of transfer functions
are all embedded in line 7 of the algorithm.

%
\begin{figure*}[t]
	\begin{algorithm}[H]
		\LinesNumbered
		\SetKwFunction{FThrMod}{ThreadModularAnalysis}
		\SetKwFunction{FFeasInterf}{GetInterfs}
		\SetKwFunction{FSeqAbs}{SeqAI}
		\SetKwProg{Fn}{Function}{:}{}
		\KwData{$ \mathtt{Tid} $ is the set of threads in the program}
		\Fn{\FThrMod{$ \mathtt{Tid} $}} {
			\tcp{Initialization}
			$ \sigma \leftarrow \phi $\;
			$ \mathcal{I} \leftarrow $ \FFeasInterf{$ \mathtt{Tid} $} \;
			\Repeat{$ \mathcal{S} = \sigma $}{
				$ \mathcal{S} \leftarrow \sigma $\;
				\ForEach{$ t \in \mathtt{Tid} $}{
					$ \sigma \leftarrow \sigma\ \squplus\ $\FSeqAbs{$ t, \mathcal{S}, \mathcal{I}(t) $}
				}
			}
		}
		\caption{TMAI}
		\label{alg:thread-modular}
	\end{algorithm}
\end{figure*}

\subsection{A Note on Precision} \label{subsec:improve}
When the older $\color{brown}{sb}$-ordered stores are forgotten in a {\tt mo} poset and those
program states having the same {\tt mo} poset are combined, it 
results in the merging of multiple program executions into a single
over-approximation.
In theory, it is possible that one or more forgotten (older) stores
were critical to prove the property. We can achieve higher precision
if we can discern such critical stores 
and preserve the ordering constraints over such stores in
the {\tt mo} posets.

In our study, we found that many benchmarks that model mutual exclusion
under the \RA memory model use {\tt rmw} instructions 
as synchronization fences.
These {\tt rmw} events are instances of critical stores, and
%
we flag  them as such
and preserve all the older {\tt rmw} instructions in $ \pomo{x} $.

\textbf{Updated Semantics of $\mathtt{RMW}$}
\label{subsec:updated-rmw}
The
semantics of $\mathtt{RMW}$
given in Figure~\ref{fig:con-sem} (for a shared variable $x$), while sound,
are not precise according to \RA semantics. We update the semantics in the
following way: 
the consistency check of two elements 
$ p_1= (\modord{M}{x}, \pomo{\modord{M}{x}}) $ and 
$ p_2=(\modord{N}{x}, \pomo{\modord{N}{x}}) $ 
will return true iff $ p_1 \iscons p_2 \andspace 
\forall \rmw_1 \in p_1, \rmw_2 \in p_2, ((\rmw_1, \rmw_2) \in 
\pomo{\modord{M}{x}} \orspace (\rmw_1, \rmw_2) \in \pomo{\modord{N}{x}} 
\orspace (\rmw_2, \rmw_1) \in \pomo{\modord{M}{x}} \orspace (\rmw_2, \rmw_1) \in 
\pomo{\modord{N}{x}} ) $.
The mentioned update prohibits the combination of those two partial 
orders such that if they were to be combined then 
the {\tt rmw} events  no longer remain in a total order.

\subsection{Loops and Termination} \label{subsec:loops}
Widening \cite{CousotCousot-92} is generally used to handle
non-terminating loops or to accelerate fix-point computation in
programs.
%
Consider a loop that contains store operations. The value to be stored
can be over-approximated using widening. Since {\tt mo} posets contain
abstracted execution histories, adding a store event in posets at
least once for each store instruction within the loop will suffice to
inform that the store has occurred at least once in the execution.
However, one can always choose to add different events corresponding
to the same store instruction depending on the precision requirement
and then widen using $\widen$, as necessary.
%

Note that one can use  widening after analyzing some fixed $n$
iterations of a program loop.  In particular, widening is applied in
the transfer function for store and rmw in function $\mathtt{SeqAI}$.

\section{Implementation and Evaluation} \label{sec:experiments}

In this section, we discuss the details of \tool's implementation and
evaluation.  In the absence of TMAI tools for \RA programs, 
we have shown the comparison of \tool with the existing tools 
designed for the \RA memory model. 
\vbmc~\cite{AbdullaKrishna-PLDI19} is the most recent  
BMC technique among these tools. Other static tools such as 
\textsc{Cppmem} and \textsc{Herd} are not designed as verification tools.
\textsc{Cppmem} is designed to help investigate possible ordering 
relations in programs under the C/C++11 memory model. It computes all the
relations of all possible executions. \textsc{Herd} is designed to
generate litmus tests for different memory models or to simulate a
memory model.  Both of these tools are relatively very slow compared to
existing verification or bug-finding tools.  We have also
compared \tool with dynamic tools such as
\textsc{CDSChecker}~\cite{Norris-TPLAS2016}, 
\textsc{Tracer}~\cite{Abdulla-OOPSLA2018}, and
\rcmc~\cite{KokologiannakisVafeiadid-POPL18} to evaluate how well
\tool performs as a refutation tool; although the input coverage guarantee of \tool
and dynamic checkers is quite different.


\subsection{Implementation}
 \tool is implemented as
an LLVM Compiler analysis pass written in C++ (code size $\sim$ 5.4KLOC).
\tool uses the \textsc{Apron} 
library \cite{Apron}  for manipulating the variable values in 
octagon and interval numerical abstract domains.
\tool takes as input an LLVM IR of an \RA program compiled with \texttt{-O1} flag,
and analyzes  user assertions in programs;
if assertions are not provided, then it can
generate the set of reachable program states at load operations
for further reasoning.  
In addition to the transfer functions in Figure~\ref{fig:con-sem},
\tool supports  $ \lock $ and $ \unlock $ operations.
\tool currently does not support dynamic thread creation
and non-integer variables. Function calls in the program are
inlined. 
%

\noindent\textbf{Handling Loops:} \tool provides support for loops in three 
ways: (i) by using the \texttt{assume} clause, (ii) by unrolling the
loops, and (iii) by a combination of \texttt{assume} clause and loop unrolling.
The {\tt assume} clause
is useful in modeling spin-wait loops in programs. 
The option of unrolling loops is used
when either the {\tt assume} clause is inadequate (such as in non-terminating
loops),
 or when we have a fixed number of iterations in the loop (such
as counting loops). 


\noindent{\bf Experimental setup:}
We have used Ubuntu 16.04 machine with Intel(R) Xeon(R) 3.60GHz CPU
and 32 GB of RAM.   The listed analysis time for each benchmark
is an average of four runs. The analysis times reported are in seconds. 

\subsection{Summary of Benchmarks}

\noindent\textbf{Benchmarks from \textsc{Tracer}:} 
The benchmarks from \textsc{Tracer}~\cite{Abdulla-OOPSLA2018}
are known to have no assertion violations.
We craft an unfenced version of the \texttt{dijkstra} benchmark 
to introduce assertion-violating behaviors in it.
%
\texttt{CO-2+2W} benchmark has
 no interferences; we  use this benchmark to distinguish
the performance of interference-based \tool  and non-interference-based 
\vbmc and \textsc{Poet}.
The benchmark \texttt{fibonacci} has a high number of load and store operations,
and is used to stress-test interference-based techniques.
%

\noindent\textbf{Benchmarks from \vbmc:}  
The benchmarks from \vbmc\cite{AbdullaKrishna-PLDI19}
are divided into two categories:
(i) the first category has benchmarks with 
assertion violations with respect to the \RA memory model, and
%
(ii) the second category consists the
 same benchmarks with appropriate fences inserted to ensure 
mutual exclusion under \RA semantics.
%

\noindent\textbf{Driver Benchmarks:} The benchmarks \texttt{ib700wdt} and 
\texttt{keybISR} are Linux device drivers taken from 
\cite{KusanoWang-FSE17,KusanoWang-FSE16,farzan_duet_2013}. We have modified 
these benchmarks to use C11 constructs. The program \texttt{ib700wdt} 
simulates multiple writers accessing a buffer and one {\em closer} that closes 
the buffer. 
The benchmark \texttt{keybISR} is an interrupt service routine
for the keyboard. 

\subsection{Observations}
\label{sec:results-analysis}


%

\begin{table}[!t]
	\centering
        	\caption{Comparison for Bug Hunting}    
	\begin{tabular}{|c|lr|lr|c|c|c|}
		\hline
		Name & \multicolumn{2}{c|}{\tool} & \multicolumn{2}{c|}{\vbmc} 
		& \multirow{2}{*}{\textsc{CDS}} & \multirow{2}{*}{\textsc{Tracer}} 
		& \multirow{2}{*}{\rcmc} \\
		\cline{2-5}
		& T  & \#It & T & VS &  &  & \\
		\hline
		peterson3 	 & 0.12 & 3 & 0.55 & 3 & 0.01 & 0.01 & 0.05 \\
		10R1W		 & 0.02 & 2 & 3.99 & 10 & 0.01 & 0.01 & 0.03 \\
		15R1W		 & 0.03 & 2 & 24.45 & 15 & 0.02 & 0.01 & 0.03 \\
		szymanski(7)     & 0.06 & 1 & 6.58 & 2 & \TO & \TO & \TO \\
		fmax(2,7)	 & 1.00 & 2 & \red{\xmark} & - & 0.15 & 0.05 & \TO \\
		\hline
	\end{tabular}
 \linebreak
        \centering
        \TO: Timeout (10 min), $\red{\times}$: Did not run
\label{tab:comp-vbmc-bug}
\end{table}

\begin{table}[!t]
        \centering
        \caption{Comparison for Proof of Correctness.}
	\begin{tabular}{|c|lr|c|c|c|c|}
		\hline
		Name & \multicolumn{2}{c|}{\tool} & {\vbmc}
		& \multirow{2}{*}{\textsc{CDS}} 
		& \multirow{2}{*}{\textsc{Tracer}} & \multirow{2}{*}{\rcmc} \\
		\cline{2-4}
		& T  & \#It & T &  &  & \\
		\hline
		CO-2+2W(5)  	  & 0.01 & 3 & 0.32 & 0.01 & 0.01 & 17.26 \\
		CO-2+2W(15)  	  & 0.02 & 3 & 1.29 & 0.02 & 0.01 &  \TO  \\
		dijkstra\_fen 	  & 0.10 & 5 & $206.70^{\red{\dag}}$ & 0.01 & 0.01 & 0.03 \\
		burns\_fen  	  & 0.02 & 4 & $37.37^{\red{\dag}}$ & 0.02 & 0.01 & 0.02 \\
		peterson\_fen	  & 0.10 & 6 & $44.12^{\red{\dag}}$ & 0.02 & 0.01 & 0.03 \\
		tbar  			  & 0.04 & 6 & 18.58 & 0.02 & 0.01 & 0.14 \\
		hehner\_c11		  & 0.03 & 6 & $107.16^{\red{\dag}}$ & 0.07 & 0.02 & 0.04 \\
		red\_co\_20		  & 0.04 & 3 & 31.47 & 23.32 & 0.13 &  \TO  \\
		exp\_bug\_6 	  & 0.45 & 6 & \red{\xmark} & 97.13 & 0.96 & 37.82 \\
		exp\_bug\_9 	  & 0.57 & 6 & \red{\xmark} & \TO  & 2.98 & 437.47 \\
		stack\_true(12)	  & 0.06 & 4 & \red{\xmark} & \TO & 589.81 & \TO \\
		ib700wdt (1)  	  & 0.01 & 3 & 31.73 & 0.01 & 0.01 & 0.02 \\
		ib700wdt (20) 	  & 0.05 & 3 & \TO & 0.01 & 0.01 & \TO \\
		ib700wdt (40) 	  & 0.07 & 3 & \TO & 0.01 & 0.01 & \TO \\
		keybISR			  & 0.01 & 4 & 0.01 & 0.01 & 0.01 & 0.03 \\
		fibonacci   	  & $0.11^{\red{\dag}}$ & 5 & 310.75 & \TO & 56.4 & 20.61 \\
		lamport\_fen 	  & $0.17^{\red{\dag}}$ & 4 & 431.40 & 0.09 & 0.03 & 0.04 \\
		\hline
	\end{tabular}
        \linebreak
        \centering
        $\red{\dag}$:False positive , \TO: Timeout (10 min) , $\red{\times}$: Did not run
	\label{tab:comp-vbmc-no-bug}
                
\end{table}



\noindent{\bf Comparison of \tool with {\sc \vbmc}:}
Tables \ref{tab:comp-vbmc-bug} and \ref{tab:comp-vbmc-no-bug} show the 
performance comparison of \tool and {\sc \vbmc} for discovering assertion violations
 and proving programs correct, respectively. 
\vbmc
with the view-bound of two,
which is the same bound used in \cite{AbdullaKrishna-PLDI19}, 
is insufficient to prove the properties in the program correct.
We increase the view bound one at a time 
and report the cumulative time.
\tool found the assertion
violations in benchmarks of Table \ref{tab:comp-vbmc-bug} in
 better time than \vbmc. It is worth noting that
in {\tt peterson3}, {\tt 10R1W}, and {\tt 15R1W}, {\sc \vbmc} could not 
find the violation with the tool's default bound of two. 

The results of \vbmc can be considered proof only 
if view bounding is relaxed and the {\em unwiding assertions} (in CBMC) hold.
However, we could not find an
option in \vbmc to disable view bounding.  
Thus, we made a decision to run \vbmc with a view-bound of 500
(assuming it to be sufficiently large) for the benchmarks in
Table \ref{tab:comp-vbmc-no-bug}.  The results in
Table \ref{tab:comp-vbmc-no-bug} illustrate that the runtimes of
\tool are consistently better than that of \vbmc. 
 \vbmc
was unable to analyze benchmarks marked with $\red{\times}$,
since they have mutex lock/unlock operations.

Many of the mutual exclusion benchmarks have
fences, which are implemented with {\tt rmw} operations. These {\tt rmw}
operations are critical in order to prove the property. As a matter of
fact,
\tool produces false positives without the improvements discussed
in \textsection\ref{subsec:improve}. Identifying {\tt rmw} operations as
critical operations and not deleting older $\color{brown}{sb}$-ordered
{\tt rmw} operations enables \tool to attain the sought precision.

\noindent {\bf False positives in \tool.}
The last two rows in Table \ref{tab:comp-vbmc-no-bug} shows the false positive results
produced by \tool. Our technique combines the  states
of different executions (having the same abstract modification order)
into a single abstracted program state. 
This results in an over-approximation of values leading to the observed
false positives in \texttt{fibonacci} and \texttt{lamport\_fen} 
benchmarks. 
%
For instance, the false positive in \texttt{lamport\_fen} is caused by
two different branch conditions (which cannot be true simultaneously
in any concrete state) evaluating to true under the abstracted program
states.
%

\noindent{\bf Comparison of \tool with dynamic tools:}
The results in Table \ref{tab:comp-vbmc-bug} indicate that \tool
performs competitively or faster than dynamic tools on these
benchmarks. Evidently, most of the executions of these benchmarks are
buggy. Hence, the probability of dynamic analyses finding the first
explored execution to be buggy is very high, leading to their
considerably fast analysis times.
The results in Table \ref{tab:comp-vbmc-no-bug} show the analysis time
over non-buggy benchmarks. 

\noindent{\bf Comparison of \tool with \textsc{Poet}:} 
%
\textsc{Poet} is unsound under the \RA model and reports
false negatives in most of the benchmarks from
Table \ref{tab:comp-vbmc-bug}.  The elapsed time when \textsc{Poet}
produced sound results is as follows: (i) TO for \textsc{Poet} on
on \texttt{10R1W} and \texttt{15R1W} while \tool analyzes them in
$\sim0.03s$, and (ii) \textsc{Poet} takes $80.43s$ seconds on \texttt{fmax(2,7)},
while \tool analyzes the benchmark in $\sim1s$.  
%


\section{Conclusions} \label{sec:conclusion}
We have presented a thread modular analysis technique for \RA programs
that uses partial orders over the set of totally ordered stores as abstract
domains. We showed that the abstract domain forms a complete lattice and
further established a  {\em Galois} correspondence between the set
of modification orders and the abstract domain.  By forgetting the
$\color{brown}{sb}$-ordered older stores, we provided a sound overapproximation on
the abstract domain, which is shown to be sound for \RA programs. We implemented
our proposal in a tool called \tool, and demonstrated its
effectiveness in not only finding bugs, but also for proving program
properties. Our experimental results revealed that \tool
attains a high degree of precision with significantly low analysis runtimes
in comparison to other tools for \RA concurrency.

\\ \ \\ 
\noindent\textbf{Acknowledgment}
We thank Sanjana Singh for her help during initial discussions. 
This work is partially supported by the Department of Science and 
Technology under the grant number DST ECR/2017/003427.

\bibliographystyle{llncs2e/splncs04}
\bibliography{section/references}

\newpage
\begin{appendix}
\section{Examples explaining \vbmc and FruitTree}\label{appendix:rel-work-details}
\subsection{}
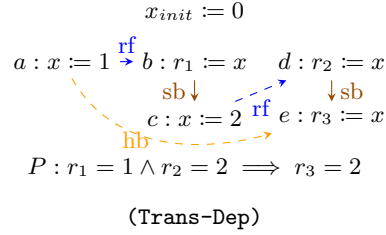
\begin{figure}[!h]
\centering
\tikzset{every picture/.style={line width=0.75pt}} 
\begin{tikzpicture}[x=1em,y=1em,yscale=1,xscale=1]
\tikzstyle{every node}=[font=\small]
\node (initx) {$ x_{init} \assign 0 $};
\node (rx1) [below =5pt of initx] {$ b: r_1 \assign x $};
\node (wx1) [left = 5pt of rx1] {$ a: x \assign 1 $};
\node (wx2) [below =8pt of rx1] {$ c: x \assign 2 $};
\node (rx2) [right =5pt of rx1] {$ d: r_2 \assign x $};
\node (rx3) [below =8pt of rx2] {$ e: r_3 \assign x $};

\draw [dashed,->,>=stealth,color=blue,thin] (wx1) -- node[midway,above] {rf} (rx1);
\draw [dashed,->,>=stealth,color=blue,thin] (wx2) -- node[midway,below] {rf} (rx2);

\draw [solid,->,>=stealth,color=brown,thin] (rx1) -- node[midway,left] {sb} (wx2);
\draw [solid,->,>=stealth,color=brown,thin] (rx2) -- node[midway,right] {sb} (rx3);

\draw [dashed,->,>=stealth,color=orange,thin] (wx1) to[out=-60,in=-165] node[left] {hb} (rx3);

\node (property) [below =5pt of wx2]{$P: r_1=1 \andspace r_2=2 \implies r_3=2$};
\node (name) [below =5pt of property] {\texttt{(Trans-Dep)}};
\end{tikzpicture}
\caption{HB relation via transitivity in (Trans-Dep)}
\label{fig:ex2-rel}
\end{figure}

Consider the example (Trans-Dep) shown in Figure~\ref{fig:ex2-rel}.
Property $ P $ in the example can be proved only if the inter-thread
dependencies are captured.  There is an ordering relation from $a$ to
$e$ when $r_1 = 1 \wedge r_2 = 2$, which is necessary to prove $P$.
{\sc FruitTree} is unable to show $P$ is valid as it does not compute
transitive inter-thread dependencies beyond two threads.

\subsection{}
Consider  the example shown in (Modified-1W2R).
\begin{figure}
\begin{center}
\begin{minipage}{0.4\textwidth}
		\small
  \begin{tabular}{c||c||c}
	\multicolumn{3}{c}{(Modified-1W2R)} \\
	$a: x \assign 1$ & $b: r_1 \assign x$ & $d: r_2 \assign x$ \\
	& $c: y  \assign  r_1$  & $e: r_3 \assign y$ \\
	\multicolumn{3}{c} {$P: r_2 \neq 1 \orspace r_3 \neq 1 $} \\
\end{tabular}
\end{minipage}
  \begin{minipage}{0.6\textwidth}
    \footnotesize
    \tikzset{every picture/.style={line width=0.75pt}} 
\begin{tikzpicture}[x=1em,y=1em,yscale=1,xscale=1]
\tikzstyle{every node}=[font=\footnotesize]
\node (initx) {$ (x_0,y_0) $};
\node (rx1) [below =5pt of initx] {$ b: (x_1,y_0) $};
\node (wx1) [left = 5pt of rx1] {$ a: (x_1,y_0) $};
\node (wy1) [below =8pt of rx1] {$ c: (x_1,y_1) $};
\node (rx2) [right =5pt of rx1] {$ d: (x_1,y_0) $};
\node (ry1) [below =8pt of rx2] {$ e: (x_1,y_1) $};
\node (mx1) [above right = 10pt and 10pt of rx2] {$m_x:(x_1,y_0)$};
\node (my1) [below =1pt of mx1] {$m_y:(x_1,y_1)$};

\draw [dashdotted,->,>=stealth,color=red] (wx1) -- (mx1);
\draw [dashdotted,->,>=stealth,color=red] (wy1) -- (my1);

\draw [dashed,->,>=stealth,color=blue] (mx1) -- (rx1);
\draw [dashed,->,>=stealth,color=blue] (mx1) -- (rx2);
\draw [dashed,->,>=stealth,color=blue] (my1) -- (ry1);



\end{tikzpicture}
\end{minipage} 
\end{center}
\caption{Program (above); View-switches (below)}
\label{fig:view-switch-ex}
\end{figure}
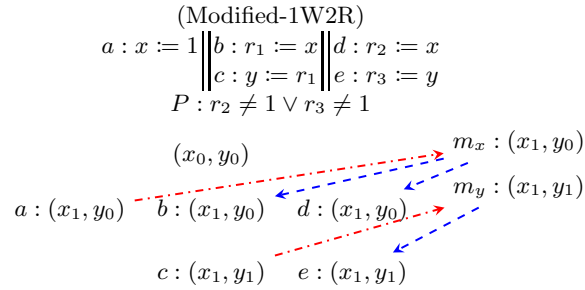

The illustration below
shows the {\em view-switches}. The pair $ (x_{i}, y_{j}) $
represents a view in which timestamps of variables $ x $ and $ y $ are
$i$ and $j $, respectively.
 The column $m_x,m_y$ represents a message-pool indicating the latest
value for $x, y$, respectively.  Each store instruction adds a message
to the message-pool. A load instruction
 reads a message from the
message-pool (blue dashed edges). The
number of dashed edges represents the
\emph{view-switch} count of the program.
When thread 1
executes the instruction \texttt{a}, it updates its view for variable $ x $, 
generating a new view $ (x_1,y_0) $, which is added to the pool $ m_x$.
The assertion can be violated only if
instruction \texttt{b} reads value 1.
Notice that with a view-switch bound set to two, \vbmc fails to
catch the violation of property $P$ in the program.
Indeed with a view-switching bound set to three can discover the
assertion violation, but adding more readers in this example
will make any fixed bound insufficient.

\section{On the Abstraction of \PO Domain} 
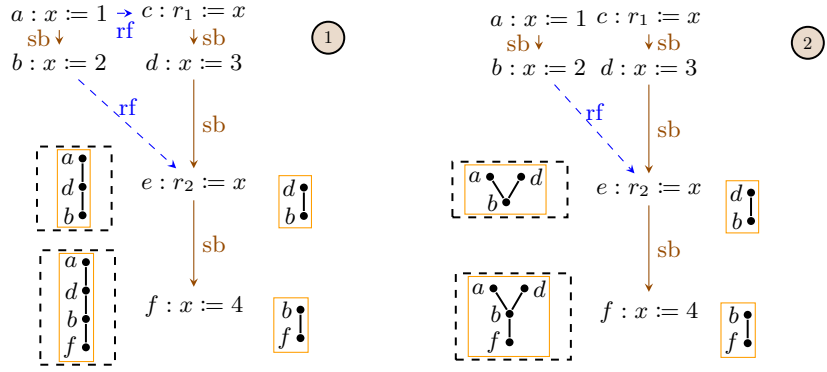
\label{appendix:abs-po}
\begin{figure}[!ht]
  \begin{minipage}{0.5\textwidth}
    \centering
    \tikzset{every picture/.style={line width=0.75pt}} 
\begin{tikzpicture}[x=1em,y=1em,yscale=-1,xscale=-1]
\tikzstyle{every node}=[font=\footnotesize]
\node (initx) [circle,draw,fill=brown!20,scale=0.75] at (-8,3) {$1$};
\node (wx1) at (3,2) {$ a: x \assign 1 $};
\node (wx2) at (3,4)   {$ b: x \assign 2 $};
\node (rx1) at (-2.5,2) {$ c: r_1 \assign x $};
\node (wx3) at (-2.5,4) {$ d: x \assign 3 $};
\node (rx2) at (-2.5,9) {$ e: r_2 \assign x $};
\node (wx4) at (-2.5,14) {$ f: x \assign 4 $};
\node (sigE) [left=28pt of rx2] {$ $};
\node (poEn1) [circle,fill=black,inner sep=0pt,minimum size=3pt, above right=5pt and 8pt of sigE] {};
\node (Ea) [left=1pt of poEn1, inner sep=0pt] {$a$};
\node (poEn2)[circle,fill=black,inner sep=0pt,minimum size=3pt, below=7pt of poEn1] {};
\node (Ed) [left=1pt of poEn2, inner sep=0pt] {$d$};
\node (poEn3)[circle,fill=black,inner sep=0pt,minimum size=3pt, below=7pt of poEn2] {};
\node (Eb) [left=1pt of poEn3, inner sep=0pt] {$b$};
\draw [color=black ] (poEn1) -- (poEn2);
\draw [color=black ] (poEn2) -- (poEn3);
\node (poE) [draw,fit=(poEn1)(Ea)(poEn2)(Eb)(poEn3)(Ed), color=orange, thin, inner sep=1pt] {};
\node (EVal) [right=0pt of poE] {$ $};
\node (sigEState) [rectangle,dashed,draw, fit=(poE)(EVal)(sigE), inner sep=1pt] {};

\node (sigF) [left=28pt of wx4] {$ $};
\node (poFn1) [circle,fill=black,inner sep=0pt,minimum size=3pt, above right=12pt and 8pt of sigF] {};
\node (Fa) [left=1pt of poFn1, inner sep=0pt] {$a$};
\node (poFn2)[circle,fill=black,inner sep=0pt,minimum size=3pt, below=7pt of poFn1] {};
\node (Fb) [left=1pt of poFn2, inner sep=0pt] {$d$};
\node (poFn3)[circle,fill=black,inner sep=0pt,minimum size=3pt, below=7pt of poFn2] {};
\node (Fd) [left=1pt of poFn3, inner sep=0pt] {$b$};
\node (poFn4)[circle,fill=black,inner sep=0pt,minimum size=3pt, below=7pt of poFn3] {};
\node (Ff) [left=1pt of poFn4, inner sep=0pt] {$f$};
\draw [color=black ] (poFn1) -- (poFn2);
\draw [color=black ] (poFn2) -- (poFn3);
\draw [color=black ] (poFn3) -- (poFn4);
\node (poF) [draw,fit=(poFn1)(Fa)(poFn2)(Fb)(poFn3)(Fd)(poFn4)(Ff), color=orange, thin, inner sep=1pt] {};
\node (FVal) [right=0pt of poF] {$ $};
\node (sigFState) [rectangle,dashed,draw, fit=(poF)(FVal)(sigF), inner sep=1pt] {};

\node (sigEAbs) [right=2pt of rx2] {$ $};
\node (poEn1Abs) [circle,fill=black,inner sep=0pt,minimum size=3pt, above right=-6pt and 8pt of sigEAbs] {};
\node (EdAbs) [left=1pt of poEn1Abs, inner sep=0pt] {$d$};
\node (poEn2Abs)[circle,fill=black,inner sep=0pt,minimum size=3pt, below=7pt of poEn1Abs] {};
\node (EbAbs) [left=1pt of poEn2Abs, inner sep=0pt] {$b$};
\draw [color=black ] (poEn1Abs) -- (poEn2Abs);
\node (poEAbs) [draw,fit=(poEn1Abs)(EdAbs)(poEn2Abs)(EbAbs), color=orange, thin, inner sep=1pt] {};
\node (EValAbs) [right=0pt of poEAbs] {$$};
\node (sigEStateAbs) [fit=(poEAbs)(sigEAbs), inner sep=1pt] {}; 

\node (sigFAbs) [right=2pt of wx4] {$ $};
\node (poFn1Abs) [circle,fill=black,inner sep=0pt,minimum size=3pt, above right=-6pt and 8pt of sigFAbs] {};
\node (FbAbs) [left=1pt of poFn1Abs, inner sep=0pt] {$b$};
\node (poFn2Abs)[circle,fill=black,inner sep=0pt,minimum size=3pt, below=7pt of poFn1Abs] {};
\node (FfAbs) [left=1pt of poFn2Abs, inner sep=0pt] {$f$};
\draw [color=black ] (poFn1Abs) -- (poFn2Abs);
\node (poFAbs) [draw,fit=(poFn1Abs)(FbAbs)(poFn2Abs)(FfAbs), color=orange, thin, inner sep=1pt] {};
\node (FValAbs) [right=0pt of poFAbs] {$ $};
\begin{scope}[on background layer]
\node (sigFStateAbs) [fit=(poFAbs)(FValAbs)(sigFAbs), inner sep=1pt] {}; 
\end{scope}

\draw [dashed,->,>=stealth,color=blue,thin] (wx1) -- node[midway,below] {rf} (rx1);
\draw [dashed,->,>=stealth,color=blue,thin] (wx2) -- node[midway,above] {rf} (rx2);
%
\draw [solid,->,>=stealth,color=brown,thin] (wx1) -- node[midway,left] {sb} (wx2);
\draw [solid,->,>=stealth,color=brown,thin] (rx1) -- node[midway,right] {sb} (wx3);
\draw [solid,->,>=stealth,color=brown,thin] (wx3) -- node[midway,right] {sb} (rx2);
\draw [solid,->,>=stealth,color=brown,thin] (rx2) -- node[midway,right] {sb} (wx4);
%

\end{tikzpicture}
  \end{minipage}
  \begin{minipage}{0.5\textwidth}
    \centering
    \tikzset{every picture/.style={line width=0.75pt}} 
\begin{tikzpicture}[x=1em,y=1em,yscale=-1,xscale=-1]
\tikzstyle{every node}=[font=\footnotesize]
\node (initx) [circle,draw,fill=brown!20,scale=0.75] at (-8,3) {$2$};
\node (wx1) at (3,2) {$ a: x \assign 1 $};
\node (wx2) at (3,4)   {$ b: x \assign 2 $};
\node (rx1) at (-1.5,2) {$ c: r_1 \assign x $};
\node (wx3) at (-1.5,4) {$ d: x \assign 3 $};
\node (rx2) at (-1.5,9) {$ e: r_2 \assign x $};
\node (wx4) at (-1.5,14) {$ f: x \assign 4 $};
\node (sigE) [left=43pt of rx2] {$ $};
\node (poEn1) [circle,fill=black,inner sep=0pt,minimum size=3pt, above right=0pt and 5pt of sigE] {};
\node (Ea) [left=1pt of poEn1, inner sep=0pt] {$a$};
\node (poEn2)[circle,fill=black,inner sep=0pt,minimum size=3pt, right=8pt of poEn1] {};
\node (Ed) [right=1pt of poEn2, inner sep=0pt] {$d$};
\node (poEn3)[circle,fill=black,inner sep=0pt,minimum size=3pt, below left=7pt and 3pt of poEn2] {};
\node (Eb) [left=1pt of poEn3, inner sep=0pt] {$b$};
\draw [color=black ] (poEn1) -- (poEn3);
\draw [color=black ] (poEn2) -- (poEn3);
\node (poE) [draw,fit=(poEn1)(Ea)(poEn2)(Eb)(poEn3)(Ed), color=orange, thin, inner sep=1pt] {};
\node (EVal) [right=0pt of poE] {$ $};
\node (sigEState) [rectangle,dashed,draw, fit=(poE)(EVal)(sigE), inner sep=1pt] {};

\node (sigF) [left=43pt of wx4] {$ $};
\node (poFn1) [circle,fill=black,inner sep=0pt,minimum size=3pt, above right=4pt and 5pt of sigF] {};
\node (Fa) [left=1pt of poFn1, inner sep=0pt] {$a$};
\node (poFn2)[circle,fill=black,inner sep=0pt,minimum size=3pt, right=8pt of poFn1] {};
\node (Fd) [right=1pt of poFn2, inner sep=0pt] {$d$};
\node (poFn3)[circle,fill=black,inner sep=0pt,minimum size=3pt, below left=7pt and 3pt of poFn2] {};
\node (Fb) [left=1pt of poFn3, inner sep=0pt] {$b$};
\node (poFn4)[circle,fill=black,inner sep=0pt,minimum size=3pt, below=7pt of poFn3] {};
\node (Ff) [left=1pt of poFn4, inner sep=0pt] {$f$};
\draw [color=black ] (poFn1) -- (poFn3);
\draw [color=black ] (poFn2) -- (poFn3);
\draw [color=black ] (poFn3) -- (poFn4);
\node (poF) [draw,fit=(poFn1)(Fa)(poFn2)(Fb)(poFn3)(Fd)(poFn4)(Ff), color=orange, thin, inner sep=1pt] {};
\node (FVal) [right=0pt of poF] {$ $};
\node (sigFState) [rectangle,dashed,draw, fit=(poF)(FVal)(sigF), inner sep=1pt] {};

\node (sigEAbs) [right=2pt of rx2] {$ $};
\node (poEn1Abs) [circle,fill=black,inner sep=0pt,minimum size=3pt, above right=-6pt and 5pt of sigEAbs] {};
\node (EdAbs) [left=1pt of poEn1Abs, inner sep=0pt] {$d$};
\node (poEn2Abs)[circle,fill=black,inner sep=0pt,minimum size=3pt, below=7pt of poEn1Abs] {};
\node (EbAbs) [left=1pt of poEn2Abs, inner sep=0pt] {$b$};
\draw [color=black ] (poEn1Abs) -- (poEn2Abs);
\node (poEAbs) [draw,fit=(poEn1Abs)(EdAbs)(poEn2Abs)(EbAbs), color=orange, thin, inner sep=1pt] {};
\node (EValAbs) [right=0pt of poEAbs] {$ $};
\node (sigEStateAbs) [fit=(poEAbs)(EValAbs)(sigEAbs), inner sep=1pt] {}; 

\node (sigFAbs) [right=2pt of wx4] {$ $};
\node (poFn1Abs) [circle,fill=black,inner sep=0pt,minimum size=3pt, above right=-6pt and 5pt of sigFAbs] {};
\node (FbAbs) [left=1pt of poFn1Abs, inner sep=0pt] {$b$};
\node (poFn2Abs)[circle,fill=black,inner sep=0pt,minimum size=3pt, below=7pt of poFn1Abs] {};
\node (FfAbs) [left=1pt of poFn2Abs, inner sep=0pt] {$f$};
\draw [color=black ] (poFn1Abs) -- (poFn2Abs);
\node (poFAbs) [draw,fit=(poFn1Abs)(FbAbs)(poFn2Abs)(FfAbs), color=orange, thin, inner sep=1pt] {};
\node (FValAbs) [right=0pt of poFAbs] {$ $};
\begin{scope}[on background layer]
\node (sigFStateAbs) [fit=(poFAbs)(FValAbs)(sigFAbs), inner sep=1pt] {};
\end{scope}

\draw [dashed,->,>=stealth,color=blue,thin] (wx2) -- node[midway,above] {rf} (rx2);
%
\draw [solid,->,>=stealth,color=brown,thin] (wx1) -- node[midway,left] {sb} (wx2);
\draw [solid,->,>=stealth,color=brown,thin] (rx1) -- node[midway,right] {sb} (wx3);
\draw [solid,->,>=stealth,color=brown,thin] (wx3) -- node[midway,right] {sb} (rx2);
\draw [solid,->,>=stealth,color=brown,thin] (rx2) -- node[midway,right] {sb} (wx4);
%

\end{tikzpicture}
  \end{minipage}
\end{figure}

Consider execution \circled{1}.  At $e$, our analysis will compute the
state with $PO_x$ as shown on the left of $e$ when specific interferences,
$\rf{a}{c}$ and $\rf{b}{e}$, are applied. 
Since $\seqb{a}{b}$, the analysis removes $a$
from the abstrct poset shown in the right of $e$. When store
$f$ is analyzed, we apply similar arguments as above
to obtain the poset (left) and its abstraction (right). 
Now consider execution 
 \circled{2} but with the application of single interference $\rf{b}{e}$.
 We observe that while $PO_x$ at $e$ and $f$ are different from
 the corressponding $PO_x$ in \circled{1},
 the abstracted $PO_x$ are the same in the two executions.

\section{{\tt mo} Losets to Posets} \label{appendix:lo-po}
\setcounter{lemma}{0}
\setcounter{theorem}{0}

\subsection{Proofs}
\begin{lemma} 
	$ (\conla, \conleq) $, is a poset.
\end{lemma}
\noindent\textbf{Reflexive:} Let $ t \defeq (S, \totordset{}{p}) $ be an element 
in $ \conla $. Since $ S \supseteq S $ and $ \forall \mo{i} \in \totordset{}{p} 
\exists \mo{j} \in \totordset{}{p}\
. \forall a,b \in S\ a \mo{i} b \implies a \mo{j} b) $ is true for all $ i = j $.
Hence the relation $ \conleq $ is reflexive.

\noindent\textbf{Transitive}: Let for some $ t_1 \defeq (S_1, \totordset{1}{m}) $,
$ t_2 \defeq (S_2, \totordset{2}{n}) $ and $ t_3 \defeq (S_3, \totordset{3}{p}) $
we have $ t_1 \conleq t_2 $ and $ t_2 \conleq t_3 $. 
From $ t_1 \conleq t_2 \iff (S_1 \supseteq S_2 \andspace \forall \mo{1_i} 
\in \totordset{1}{m} \exists \mo{2_j} \in \totordset{2}{n}\
. \forall a,b \in S_2\ a \mo{1_i} b \implies a \mo{2_j} b) $ and from $ t_2 
\conleq t_3 \iff (S_2 \supseteq
S_3 \andspace \forall \mo{2_i} \in \totordset{2}{n} \exists \mo{3_j} 
\in \totordset{3}{p}\
. \forall a,b \in S_3\ a \mo{2_i} b \implies a \mo{3_j} b) $.
Now we have $ S_1 \supseteq S_2$ and $ S_2 \supseteq S_3 $. 
Hence, $ S_1 \supseteq S_3 $.
Similarly, we have $ \forall \mo{1_i} 
\in \totordset{1}{m} \exists \mo{2_j} \in \totordset{2}{n}\
. \forall a,b \in S_2\ a \mo{1_i} b \implies a \mo{2_j} b $ and 
$ \forall \mo{2_i} \in \totordset{2}{n} \exists \mo{3_j} 
\in \totordset{3}{p}\ . \forall a,b \in S_3\ a \mo{2_i} b 
\implies a \mo{3_j} b $. Hence we have $ \forall \mo{1_i} \in 
\totordset{1}{m} \exists \mo{3_j} \in \totordset{3}{p}\
. \forall a,b \in S_3\ a \mo{1_i} b \implies a \mo{3_j} b $,
{\em i.e.}, $t_1 \conleq t_3$.
Hence $ \conleq $ is a transitive relation.

\noindent\textbf{Anti-symmetric:} Let for some $ t_1 \defeq 
(S_1, \totordset{1}{m}) $ and $ t_2 \defeq (S_2, \totordset{2}{n}) $, 
$ t_1 \conleq t_2 $ and $ t_2 \conleq t_1 $. 
We know that $ t_1 \conleq t_2 \iff 
(S_1 \supseteq S_2 \andspace \forall \mo{1_i} 
\in \totordset{1}{m} \exists \mo{2_j} \in \totordset{2}{n}\
. \forall a,b \in S_2\ a \mo{1_i} b \implies a \mo{2_j} b) $.
Now $ t_2 \conleq t_1 \iff S_2 \supseteq S_1 $. Hence we 
have $ S_1 = S_2 $. Now we can write $ \forall \mo{1_i} 
\in \totordset{1}{m} \exists \mo{2_j} \in \totordset{2}{n}\
. \forall a,b \in S_2\ a \mo{1_i} b \implies a \mo{2_j} b $ as 
$ \totordset{1}{m} \subseteq \totordset{2}{n} $. Similarly, 
from $ t_2 \conleq t_1 $, we get $ \totordset{2}{n} \subseteq 
\totordset{1}{m} $. Which is possible only if $ \totordset{1}{m} 
= \totordset{2}{n} $. Hence $ t_1 = t_2 $.

The relation $ \conleq $ is transitive, reflexive and 
anti-symmetric by definition. 
Hence $ (\conla, \conleq) $ forms a poset.
\qed

\begin{lemma} 
The operators $ \join $ and $\meet$ defines lub and glb of any two
elements of $\mathcal{P}$, respectively. 
\end{lemma}
\textbf{Proof of lub:}
	Let for any two elements $ p_1=(\modord{M}{x}, \pomo{1})$ and 
	$ p_2=(\modord{N}{x}, \pomo{2}) $ from set $ \mathcal{P} $, 
	$ p = p_1 \join p_2 $. Let $ p = (\modord{Q}{x}, \pomo{}) $.
	Let $ p_a = (\modord{A}{x}, \pomo{a}) \in \mathcal{P} $ be some other upper bound of the elements 
	$ p_1 $ and $ p_2 $, i.e, $ p_1 \leqlattice p_a \andspace p_2 \leqlattice p_a $. 
	By definition, $ p_1 \leqlattice p_a \iff \modord{M}{x} \supseteq \modord{A}{x} 
	\andspace \inpomo{a}{e_1}{e_2} \implies \inpomo{1}{e_1}{e_2} $ 
	and $ p_2 \leqlattice p_a \iff \modord{N}{x} \supseteq \modord{A}{x} \andspace 
	\inpomo{a}{e_1}{e_2} \implies \inpomo{2}{e_1}{e_2} $
	By definition of $ \join $ operator, $ \modord{Q}{x} = \modord{M}{x} \cap 
	\modord{N}{x} $ 	and $ \pomo{} = \pomo{1} \cap 
	\pomo{2} $.
	Since $ p_1 \leqlattice p_a \andspace p_2 \leqlattice p_a $, 
	$ (\modord{M}{x} \supseteq \modord{A}{x} \andspace \modord{N}{x} \supseteq \modord{A}{x}) 
	\andspace (\inpomo{a}{e_1}{e_2} \implies \inpomo{1}{e_1}{e_2} 
	\andspace \inpomo{2}{e_1}{e_2}) $. Hence, $ (\modord{M}{x} \cap \modord{N}{x} 
	\supseteq \modord{A}{x}) \andspace ((e_1,e_2) \in \inpomo{a}{}{} \implies 
	(e_1,e_2) \in \inpomo{1}{}{} \cap \inpomo{2}{}{}) $.
	Hence $ \modord{Q}{x} \supseteq \modord{A}{x} \andspace \inpomo{a}{e_1}{e_2} 
	\implies \inpomo{}{e_1}{e_2} $. Therefore, $ p \leqlattice p_a $. 
	Hence $ p_1 \join p_2 $ is the \emph{lowest upper bound} of $ p_1 $ and $ p_2 $.

\noindent\textbf{Proof of glb:}
	 Let for any two elements $ p_1=(\modord{M}{x}, \pomo{1})$ and 
	 $ p_2=(\modord{N}{x}, \pomo{2}) $ from set $ \mathcal{P} $, 
	 $ p = p_1 \meet p_2 $.
	 Let $ p = (\modord{Q}{x}, \pomo{}) $.
	 Let $ p_a = (\modord{A}{x}, \pomo{a}) \in \mathcal{P} $ be some other lower bound of the elements 
	 $ p_1 $ and $ p_2 $, i.e, $ p_a \leqlattice p_1 \andspace p_a \leqlattice p_2 $. 
	 By definition, $ p_a \leqlattice p_1 \iff \modord{A}{x} \supseteq \modord{M}{x}
	 \andspace \inpomo{1}{e_1}{e_2} \implies \inpomo{a}{e_1}{e_2} $ 
	 and $ p_a \leqlattice p_2 \iff \modord{A}{x} \supseteq \modord{N}{x} \andspace 
	 \inpomo{2}{e_1}{e_2} \implies \inpomo{a}{e_1}{e_2} $.
	 
	 If $ p_1 \iscons p_2 $, by the definition of meet operation $ \modord{Q}{x} = 
	 \modord{M}{x} \cup \modord{N}{x} $ and $ \pomo{} = \pomo{1} 
	 \cup \pomo{2} $.
	 Since $ \modord{A}{x} \supseteq \modord{M}{x} $ and $ \modord{A}{x} \supseteq \modord{N}{x} $,
	 then $ \modord{A}{x} \supseteq \modord{M}{x} \cup \modord{N}{x} $. 
	 Hence $ \modord{A}{x} \supseteq \modord{Q}{x} $.
	 From $ \inpomo{1}{e_1}{e_2} \implies \inpomo{a}{e_1}{e_2} $
	 and $ \inpomo{2}{e_1}{e_2} \implies \inpomo{a}{e_1}{e_2} $,
	 we have $ \inpomo{1}{e_1}{e_2} \orspace \inpomo{2}{e_1}{e_2}
	 \implies \inpomo{a}{e_1}{e_2} $. From definition of meet operator,
	 $ \inpomo{}{e_1}{e_2} \implies \inpomo{1}{e_1}{e_2} \orspace 
	 \inpomo{2}{e_1}{e_2} $. Hence $ \inpomo{}{e_1}{e_2} \implies 
	 \inpomo{a}{e_1}{e_2} $.
	 Therefore, $  \modord{A}{x} \supseteq \modord{Q}{x} \andspace 
	 \inpomo{}{e_1}{e_2} \implies \inpomo{a}{e_1}{e_2} $
	 Hence $ p_a \leqlattice p $ i.e., $ p $ is \emph{greatest lower bound} of $ p_1 $ and $ p_2 $.
	 
	 If $ p_1 \iscons p_2 $ is false, the only consistent partial order $ p_a $ such that 
	 $  p_a \leqlattice p_1 \andspace p_a \leqlattice p_2 $, is $ \bot $. Hence 
	 $ p_1 \meet p_2 $ gives the \emph{greatest lower bound}.
\qed

\begin{lemma} 
	$ (\mathcal{P}, \leqlattice, \join, \meet, \bot, \top) $ is a complete
	lattice, where $ \mathcal{P} $ is set of all possible partial orders
	over elements of set $ \mathtt{St} $ and $ \top $ is defined as
	empty poset and $ \bot $ is a special elements in $ \mathcal{P} $
	that is ordered below all the elements of $ \mathcal{P} $ in $
	\leqlattice $.
\end{lemma}
The relation $ \leqlattice $ is transitive, reflexive and
anti-symmetric by definition. Hence $ (\mathcal{P}, \leqlattice) $ is
a poset.  Lemma \ref{lem:lub-glb-po}  proves that the
operators $ \join $ and $ \meet $ compute a $lub$
and $glb$, respectively, of any two elements of
$ \mathcal{P} $. Hence, $ (\mathcal{P}, \leqlattice, \join, \meet, \bot, \top) $ 
is a lattice. To prove that it is a complete lattice, we need to prove 
that for every $ P \subset \mathcal{P} $, $\join$ and $\meet$ over set $P$ exits.

\noindent\textbf{$ \join $ operation over all subsets:}
Using the definition of $ \join $ operation, for some $ P \subset \mathcal{P} $, $ \bigsqcup P $ 
can be defined as $ p = (\modord{Q}{x}, \pomo{}) $, where 
$\modord{Q}{x} = \intersectionof{p_i \in P }{} \modord{M(i)}{x}$ 
and $\pomo{} = \intersectionof{p_i \in P }{} \pomo{i} $, where $ p_i=(\modord{M(i)}{x}, \pomo{i}) $.
Let $ p_a = (\modord{A}{x}, \pomo{a}) \in \mathcal{P} $ be an upper bound of $ P $.
We know $ p_a $ exists because by definition, $ \top $ is an upper bound of all the lattice elements. 
Since $ p_a $ is an upper bound of $ P $, $ \forall p_i \in P $ 
$ p_i \leqlattice p_a \iff \modord{M(i)}{x} \supseteq \modord{A}{x} \andspace 
(\inpomo{a}{e_1}{e_2} \implies \inpomo{i}{e_1}{e_2}) $ by definition of $ \leqlattice $ operation.
Using simple set operations, we can say that
$ (\forall p_i \in P \ \modord{M(i)}{x} \supseteq \modord{A}{x}) \implies 
\intersectionof{p_i \in P }{} \modord{M(i)}{x} \supseteq \modord{A}{x} \implies 
\modord{Q}{x} \supseteq \modord{A}{x} $.
Similarly, $ (\forall p_i \in P \ (e_1, e_2) \in \pomo{a} \implies (e_1, e_2) \in \pomo{i}) \iff 
((e_1, e_2) \in \pomo{a} \implies (\forall p_i \in P \ (e_1, e_2) \in \pomo{i})) \iff 
((e_1, e_2) \in \pomo{a} \implies (e_1, e_2) \in \intersectionof{p_i \in P}{} \pomo{i}) \iff 
((e_1, e_2) \in \pomo{a} \implies (e_1, e_2) \in \pomo{}) $. Hence $ p \leqlattice p_a $.

\noindent\textbf{$ \meet $ operation over all subsets:}
Let for some $ P \subset \mathcal{P} $, $ \meet $ over set $ P $ 
can be defined as $ p = (\modord{Q}{x}, \pomo{}) $
if $ \iscons P $, then $ \modord{Q}{x} = \unionof{p_i \in P}{} \modord{M(i)}{x} $ and 
$ \pomo{} = \unionof{p_i \in P}{} \pomo{i} $ else $ \bot $. 
If $ \iscons P $ is false, the only possible lower bound of $ P $ is $ \bot $.
Hence, we need to prove the existence of glb only if  $ \iscons P $.
Let $ p_a = (\modord{A}{x}, \pomo{a}) \in \mathcal{P} $ be a lower bound of $ P $.
We know $ p_a $ exists because by definition, $ \bot $ is a lower bound 
of all the lattice elements.
Since $ p_a $ is a lower bound of $ P $, $ \forall p_i \in P $ 
$ \forall p_i \in P, p_a \leqlattice p_i $. Therefore, 
$ (\forall p_i \in P,\ \modord{A}{x} \supseteq \modord{M(i)}{x}) \iff  
\modord{A}{x} \supseteq \unionof{p_i \in P}{} \modord{M(i)}{x} \iff 
\modord{A}{x} \supseteq \modord{Q}{x} $. 
Similarly, from $ \forall p_i \in P, p_a \leqlattice p_i $, we have
$ (\forall p_i \in P, (e_1, e_2) \in \pomo{i} \implies (e_1, e_2) \in \pomo{a}) \iff 
((e_1, e_2) \in \unionof{p_i \in P}{} \pomo{i} \implies (e_1, e_2) \in \pomo{a}) \iff 
((e_1, e_2) \in \pomo{} \implies (e_1, e_2) \in \pomo{a}) $.
Now we have $ \modord{A}{x} \supseteq \modord{Q}{x} \andspace 
((e_1, e_2) \in \pomo{} \implies (e_1, e_2) \in \pomo{a}) $. Hence,
$ p_a \leqlattice p $, which proves that $ p $ is glb of set $ P $.

Therefore $ (\mathcal{P}, \leqlattice, \join, \meet, \bot, \top) $ is a complete
lattice.
%
\qed

\begin{lemma} 
	The operation $ \widen $ defines widening operator over elements of lattice 
	
\end{lemma}
The operation $ \nabla $ is widening operator over lattice 
$ (\mathcal{P}, \leqlattice, \join, \meet, \bot, \top) $ iff 
\begin{enumerate}[(i)]
	\item $ \widen $ is an upper-bound operator i.e, $ p_1, p_2 \leqlattice p_1 \widen p_2 $, and
	\item for all ascending chains $ p_0, p_1, p_2, \dots $ over elements of lattice $ \mathcal{P} $, the ascending 
	chain $ p_0^{\widen}, p_1^{\widen}, p_2^{\widen} \dots $ eventually stabilizes, where 
	$ p_i^{\widen} $ is defined as $ p_0^{\widen} = p_0 $ and $ \forall i>0, p_i^{\widen} = p_{i-1}^{\widen} \widen p_i $.
\end{enumerate}

\noindent\textbf{Proof of (i)} Let $ p = p_1 \widen p_2 $ for some $ p_1 = (\modord{M}{x}, \pomo{1}) $, 
$ p_2 = (\modord{N}{x}, \pomo{2}) $ and $ p = (\modord{Q}{x}, \pomo{}) $.
By definition of $ \widen $, $ \modord{M}{x} \supseteq \modord{Q}{x} $ and $ \modord{N}{x} \supseteq \modord{Q}{x} $.
Similarly, $ \inpomo{}{a}{b} \implies \inpomo{1}{a}{b} \andspace \inpomo{2}{a}{b} $. 
Therefore, $ p_1, p_2 \leqlattice p $. Hence $ \widen $ is an upper bound operator.

\noindent\textbf{Proof of (ii)}
Since the number of program instructions are finite in any program, we have a finite set of 
instruction labels. By definition of $ \widen $, for any $ p = (\modord{Q}{x}, \pomo{}) $ such that 
$ p = p_1 \widen p_2 $, the set $ \modord{Q}{x} $ may contain at most one event for each instruction label.
Hence in any such $ p $, the set of events $ \modord{Q}{x} $ is finite. There are only finitely many 
possible pair may exist over a finite set $ \modord{Q}{x} $. Hence, the set $ \pomo{} $ is also finite.

Therefore, we can say that in chain $ p_0^{\widen}, p_1^{\widen}, p_2^{\widen} \dots $, where 
$ p_i^{\widen} = (\modord{M(i^{\widen})}{x}, \pomo{i^{\widen}} ) $, $ \modord{M(i^{\widen})}{x} $ and 
$ \pomo{i^{\widen}} $ are finite for all $ i>0 $.
We know that, the chain $ p_0^{\widen}, p_1^{\widen}, p_2^{\widen} \dots $ is an ascending chain 
(property of upper-bound operator \cite{Nielson-2010}). Hence, from definition of $ \leqlattice $
we have $ \forall i>0, \modord{M(i^{\widen})}{x}  \subseteq \modord{M(i-1^{\widen})}{x} \andspace 
\pomo{i^{\widen}} \subseteq \pomo{i-1^{\widen}} $.
Let chain $ p_0^{\widen}, p_1^{\widen}, p_2^{\widen} \dots $ does not stabilize. Hence 
$ \forall i>0$, either $ \modord{M(i^{\widen})}{x}  \subset \modord{M(i-1^{\widen})}{x} $ or 
$ \pomo{i^{\widen}} \subset \pomo{i-1^{\widen}} $. 
We can also say that $ \forall i>0 $ either $ |\modord{M(i^{\widen})}{x}| < |\modord{M(i-1^{\widen})}{x}| $ or 
$ |\mathrm{\pomo{i^{\widen}}}| < |\mathrm{\pomo{i-1^{\widen}}}| $ where 
$ |\modord{M(i^{\widen})}{x}|, |\modord{M(i-1^{\widen})}{x}|, |\mathrm{\pomo{i^{\widen}}}| $ and 
$ |\mathrm{\pomo{i-1^{\widen}}}| $ are some natural number $ < \omega $. 
Now, $ |\modord{M(i^{\widen})}{x}| < |\modord{M(i-1^{\widen})}{x}| $ is not possible 
infinitely often since any strictly decreasing chain over $ |\modord{M(i^{\widen})}{x}| $ 
starting from some natural number $ < \omega $ will 
eventually reach minimal element 0 and can not decrease further.
Similarly, $ |\mathrm{\pomo{i^{\widen}}}| < |\mathrm{\pomo{i-1^{\widen}}}| $ is not possible infinitely often. 
Hence a strictly decreasing chain of $ (|\modord{M(i^{\widen})}{x}|,|\mathrm{\pomo{i^{\widen}}}|) $ 
form cannot be infinite. 
It means that a infinite strictly ascending chain over $ p_i^{\widen} $ is not possible.
Therefore, the chain $ p_0^{\widen}, p_1^{\widen}, p_2^{\widen} \dots $ eventually stabilizes.
\qed

\begin{lemma} \label{lem:beta-subset-apdx}
	$ (p_1, p_2) \in \beta \implies p_1 \leqlattice p_2 $.
\end{lemma}

The property is trivially true for $ p_1 = \bot $ since $ \bot $ is the
least element in lattice. For $p_1 \neq \bot$, from the definition of $\beta$,
$\modord{M}{x} \supseteq \modord{N}{x} \andspace
\pomo{2} \subseteq \pomo{1}$.
But $ \pomo{2} \subseteq \pomo{1} \iff
\inpomo{2}{a}{b} \implies \inpomo{1}{a}{b}$,
Hence, by definition of $ \leqlattice $, $ p_1 \leqlattice p_2 $.

\begin{lemma} \label{lem:greater-in-beta-apdx}
	Abstract soundness assumption holds under $ \beta $, i.e., $ \forall p, 
	p_1, p_2 \in \mathcal{P}. $  $(p, p_1) \in \beta \andspace p_1 
	\leqlattice p_2 \implies (p, p_2) \in \beta $
\end{lemma}
By Lemma \ref{lem:beta-subset}, we know that $ (p,p_1) \in \beta \implies p 
\leqlattice p_1 $. If $ p_1 = \bot $, $ p \leqlattice p_1 $ is possible if 
and only if $ p = \bot $. By definition of $ \beta $, $ \forall p_2 \in 
\mathcal{P} $, $ (\bot, p_2) \in \beta $. Now for $ p \neq \bot $,
by definition, $ (p, p_1) \in \beta \iff (\modord{M}{x} \subseteq \modord{Q}{x} 
\setminus \{a \mid \exists b \in \modord{Q}{x} \ .\ \seqb{a}{b} 
\andspace a \neq b \} \andspace \pomo{1} \subseteq 
\pomo{}) $. From $ p_1 \leqlattice p_2 $, it follows that 
$ \modord{N}{x} \subseteq \modord{M}{x} \andspace \pomo{2} 
\subseteq \pomo{1} $. Hence, we have $ \modord{N}{x} \subseteq 
\modord{M}{x} \subseteq \modord{Q}{x} \setminus \{a \mid \exists b \in 
\modord{Q}{x} \ .\ \seqb{a}{b} \andspace a \neq b \} $ and 
$ \pomo{2} \subseteq \pomo{1} \subseteq 
\pomo{} $. Hence $ (p, p_2) \in \beta $

\begin{lemma} \label{lem: linearization}
  For some set of orderings $ O_1 $ and $ O_2 $ over elements from set
  $ S $, $ O_1 \subseteq O_2 \iff L(S, O_1) \supseteq L(S, O_2) $.
\end{lemma}
	$ (\implies) $ Let for some set $ S $ and ordering relations $ O_1 \subseteq O_2 $, $ L(S, O_1) \nsupseteq L(S, O_2) $. 
	Without loss of generality, let $ \mo{} $ be an ordering relation such that $ \mo{} \notin L(S, O_1) \andspace$ $\mo{} \in L(S, O_2) $. 
	By definition, $ \mo{} $ must satisfy all the ordering defined in $ O_2 $. 
	Since $ O_1 \subseteq O_2 $, $ \mo{} $ also satisfies all the ordering in $ O_1 $. 
	Hence $ \mo{} \in L(S, O_1) $. Therefore, $ L(S, O_1) \supseteq L(S, O_2) $.
	
	$ (\impliedby) $ Let us assume that for some $ L(S, O_1) \supseteq L(S, O_2) $, $ O_1 \nsubseteq O_2 $. It means all $ \mo{} \in L(S, O_2) $ also satisfies ordering relations in $ O_1 $. Hence $ O_1 \subseteq O_2 $.
\qed

\begin{theorem}
$       (\mathcal{T}, \conleq) \galois{\alpha}{\gamma}                   
(\mathcal{P}, \leqlattice, \join, \meet, \bot, \top) $.
\end{theorem}
	Let $ t_1, t_2 \in \mathcal{T} $,and $ p_1, p_2 \in \mathcal{P} $, for some 
	$ t_1=(S_1,\totordset{1}{n}), t_2=(S_2,\totordset{2}{m}), 
	p_1=(\modord{M}{x}, \pomo{1}) $ and $ p_2=(\modord{N}{x}, \pomo{2}) $.
	To prove that $ \alpha $ is monotonic, let $ \alpha(t_1)=p_1, \alpha(t_2)=p_2 $, and $ t_2 \conleq t_1 $ without loss of generality.
	\begin{align}
		\label{eq:p1-is-alpha-t1}
		p_1 = \alpha(t_1) \iff & \modord{M}{x} = S_1 \andspace \pomo{1} =
                \intersectionof{i=1}{n} \mo{1_i} & \reason{By definition of $ \alpha $} \\ 
		\label{eq:p2-is-alpha-t2}
		p_2 = \alpha(t_2) \iff & \modord{N}{x} = S_2 \andspace \pomo{2} = \intersectionof{i=1}{m} \mo{2_i} & \reason{By definition of $ \alpha $} 
	\end{align}
	\begin{align*}
		t_2 \conleq t_1 \iff & (S_2 \supseteq S_1 \andspace 
		\forall \mo{2_i} \in \totordset{2}{m} \exists \mo{1_j} \in \totordset{1}{n}\ . \forall a,b \in S_1\ a \mo{2_i} b \implies \\
		& a \mo{1_j} b) \tag{By definition of $ \conleq $} \\
		\iff & (\modord{N}{x} \supseteq \modord{M}{x} \andspace 
		\forall \mo{2_i} \in \totordset{2}{m} \exists \mo{1_j} \in \totordset{1}{n}\ . \forall a,b \in S_1\ a \mo{2_i} b \implies  \\
		& a \mo{1_j} b) \tag{By eq \ref{eq:p1-is-alpha-t1} \& \ref{eq:p2-is-alpha-t2}} \\
		\iff & (\modord{N}{x} \supseteq \modord{M}{x} \andspace \intersectionof{i=1}{n} \mo{1_i} 
		\subseteq \intersectionof{i=1}{m} \mo{2_i}) 
		\tag{By Lemma \ref{lem: linearization}} \\
		\iff & (\modord{N}{x} \subseteq \modord{M}{x} \andspace \forall a,b \in \modord{M}{x} \ .\  \inpomo{1}{a}{b} \implies \inpomo{2}{a}{b}) \\
		\iff & (p_2 \leqlattice p_1) \tag{By definition of $ \leqlattice $} 
	\end{align*}
	Hence $ \alpha $ is monotonic.
	To prove that $ \gamma $ is monotonic, let $ \gamma(p_1)=t_1, \gamma(p_2)=t_2 $, and $ p_2 \leqlattice p_1 $ without loss of generality.
	%
	\begin{align}
		\label{eq:t1-is-gamma-p1}
		t_1=\gamma(p_1) \iff & S = \modord{Q}{x} \andspace \totordset{1}{n} = L(S_1, \pomo{1}) & \reason{By definition of $ \gamma $} \\
		\label{eq:t2-is-gamma-p2}
		t_2=\gamma(p_2) \iff & S = \modord{Q}{x} \andspace \totordset{2}{n} = L(S_2, \pomo{2}) & \reason{By definition of $ \gamma $}
	\end{align}
	\begin{align*}
		p_2 \leqlattice p_1 \iff &
		(\modord{N}{x} \supseteq \modord{M}{x} \andspace \pomo{1} \subseteq \pomo{2}) 
		\tag{By definition of $ \leqlattice $}\\
		\iff & (S_2 \supseteq S_1 \andspace \pomo{1} \subseteq \pomo{2}) 
		\tag{By eq \ref{eq:t1-is-gamma-p1} \& \ref{eq:t2-is-gamma-p2}} \\
		\iff & (S_2 \supseteq S_1 \andspace L(S_1, \pomo{1}) \supseteq L(S_1, \pomo{2})) \tag{By lemma \ref{lem: linearization}} \\
		\iff & (S_2 \supseteq S_1 \andspace \forall \mo{2_i} \in \totordset{2}{m} \exists \mo{1_j}\in \totordset{1}{n}\ . \forall a,b \in S_1\ a \mo{2_i} b \implies \\
		& a \mo{1_j} b) \\
		\iff & t_2 \conleq t_1 \tag{By definition of $ \conleq $}
	\end{align*}
	Hence $ \gamma $ is monotonic.
	Let $ \alpha(t_1)=p_1, \gamma(p_2)=t_2 $. If we can prove that $ p_2 \leqlattice p_1 \iff t_2 \conleq t_1 $, 
	$ (\alpha, \gamma) $ forms a Galois connection between the lattices. 
	\begin{align*}
		p_2 \leqlattice p_1 \iff & (\modord{N}{x} \supseteq \modord{M}{x} \andspace \inpomo{1}{a}{b} \implies \inpomo{2}{a}{b}) 
		\tag{By definition of $ \leqlattice $} \\
		\iff & (\modord{N}{x} \supseteq \modord{M}{x} \andspace \pomo{1} \subseteq \pomo{2}) \\
		\iff & (\modord{N}{x} \supseteq S_1 \andspace \intersectionof{i=1}{n} \mo{1_i} \subseteq \pomo{2}) 
		\tag{By eq \ref{eq:p1-is-alpha-t1}} \\
		\iff & (S_2 \supseteq S_1 \andspace \intersectionof{i=1}{n} \mo{1_i} \subseteq \pomo{2})
		\tag{By eq \ref{eq:t2-is-gamma-p2}} \\
		\iff & (S_2 \supseteq S_1 \andspace L(S_1,\intersectionof{i=1}{n} \mo{1_i}) \supseteq L(S_2, \pomo{2}))
		\tag{By lemma \ref{lem: linearization}} \\
		\iff & (S_2 \supseteq S_1 \andspace \forall \mo{2_i} \in \totordset{2}{m} \exists \mo{1_j} \in \totordset{1}{n}\ . \forall a,b \in S_1\ a \mo{2_i} b \implies \\ & a \mo{1_j} b
		\tag{By well-formedness of $ t_1 $ and eq \ref{eq:t2-is-gamma-p2}} \\
		\iff & t_2 \conleq t_1 \tag{By definition of $ \conleq $}
	\end{align*}
\qed

\begin{theorem} 
	There is a Galois connection among elements of 
	$ \prod_{x \in \vars} \mathcal{P}_x $ and 
	$ \prod_{x \in \vars} \mathcal{T}_x $.
\end{theorem}
A Galois connection between two posets can be lifted to a Galois 
connection in the Cartesian product of the posets \cite{Nielson-2010}, i.e.,
if $ L_1 \galois{\alpha_1}{\gamma_1} M_1 $ and $ L_2 
\galois{\alpha_2}{\gamma_2} M_2 $, then $ L_1 \times L_2 
\galois{\alpha}{\gamma} M_1 \times M_2 $, where $ \alpha(l_1, l_2) \defeq 
(\alpha_1(l_1), \alpha_2(l_2)) $, $ \gamma(m_1, m_2) \defeq (\gamma_1(m_1), 
\gamma_2(m_2)) $. We use this result to establish that the 
tuples of elements from $ \prod_{x \in \vars} \mathcal{P}_x $ in the program 
forms a lattice. Further, such a lattice will have Galois connection with 
the poset of $ \prod_{x \in \vars} \mathcal{T}_x $.
\qed

\begin{lemma} \label{lem:all-abs-beta}
	Every concrete property has an abstraction under soundness relation 
	$\beta$, i.e, $ \forall p \in \mathcal{P}, \exists p' \in \mathcal{P} 
	\ .\ (p,p') \in \beta $.
\end{lemma}
If $ p = \bot $, then $ (p,p') \in \beta $ for all $ p' \in \mathcal{P} $.
Hence the property is trivially true for $ \bot $. For  $ p \neq \bot $,
we know that $ \emptyset \subseteq \modord{Q}{x} \setminus \{a \mid \exists b \in 
\modord{Q}{x} \ .\ \seqb{a}{b} \andspace a \neq b \} $. Similarly, $ \emptyset 
\subseteq \pomo{} $. Hence, there exist an element $ (\emptyset, \emptyset) 
\in \mathcal{P} $ (i.e., $ \top $) such that $ (p, \top) \in \beta $. 
\qed
\ \\
The significance of Lemma \ref{lem:all-abs-beta} rests in the fact that every element 
in $ \mathcal{P} $ has at least one over-approximation under the soundness relation $ \beta $.

\begin{theorem} \label{thm:abspo-sound-apdx}
	Abstraction relation $ \abs{\alpha} $ is minimal sound abstraction under 
	soundness relation $ \beta $, i.e., $ (p_1, p_2) \in \beta \iff 
	\abs{\alpha}(p_1) \leqlattice p_2 $.
\end{theorem}
$(\implies)$ For $ p_1 = \bot $, $ \abs{\alpha}(p_1) = \bot $, which is the least element 
in the lattice. Let us now focus on $ p_1 \neq \bot $.
Let $ \abs{\alpha}(p_1) = p $. By definition, $ \modord{Q}{x} = \modord{M}{x} 
\setminus \{a \mid \exists b \in \modord{Q}{x}\ .\ \seqb{a}{b} \andspace a \neq b \} $. 
From $ (p_1, p_2) \in \beta $, we have $ \modord{N}{x} \subseteq 
\modord{M}{x} \setminus \{a \mid \exists b \in \modord{M}{x} \ .\ \seqb{a}{b} 
\andspace a \neq b \} $. Hence $ \modord{N}{x} \subseteq \modord{Q}{x} $. 
Let $ R = \{(a,b) \mid (a,b) \in  
\pomo{1} \andspace (a \notin \modord{Q}{x} \orspace b \notin 
\modord{Q}{x}) \} $. By definition of $ \abs{\alpha} $, we have $ 
\abs{\alpha}(p_1) = p \iff \pomo{} = \pomo{1} \setminus 
R $. By definition of $ \beta $, we have, $ (p_1, p_2) \in \beta \iff 
\pomo{2} \subseteq \pomo{1} $. 
To prove that $ p \leqlattice p_2 $, we need to prove that 
$ \pomo{2} \subseteq \pomo{} $.
Let us assume that $ (a',b') \notin \pomo{} $, $ (a',b') \in 
\pomo{1} $.
Since $ \pomo{} = \pomo{1} 
\setminus R $,  $ (a',b') \notin \pomo{} \implies (a',b') \in 
R $, which is possible only if $ a' \notin \modord{Q}{x} \orspace b' \notin 
\modord{Q}{x} $. This is true only if either $ a' \in  \{a \mid \exists b 
\in \modord{Q}{x}\ .\ \seqb{a}{b} \andspace a \neq b \} $ or $ b' \in  
\{a \mid \exists b \in \modord{Q}{x}\ .\ \seqb{a}{b} \andspace a \neq b \} $. 
Hence, $ a' \notin \modord{N}{x} $ or $ b' \notin \modord{N}{x} $, which means 
$ (a',b') \notin \pomo{2} $. Therefore, $ p \leqlattice p_2 $.

$(\impliedby)$ By defintion of $\beta$, for $p_1 = \bot, \forall p_2 \in 
\mathcal{P} \ (p1,p_2) \in \beta$. Hence it is trivially true. For 
$p_1 \neq \bot$, $\abs{\alpha}(p_1) = p \iff \modord{Q}{x} = \modord{M}{x} 
\setminus \{a \mid \exists b \in \modord{M}{x}\ .\ 
\seqb{a}{b} \andspace a \neq b \} $. Hence $ \modord{Q}{x} \subseteq 
\modord{M}{x} \setminus \{a \mid \exists b \in \modord{M}{x}\ .\ 
\seqb{a}{b} \andspace a \neq b \} $. Similarly, $\pomo{} = \pomo{1} 
\setminus \{(a,b) \mid (a,b) \in  
\pomo{} \andspace (a \notin \modord{Q}{x} \orspace b \notin 
\modord{Q}{x}) \}$. Hence $\pomo{} \subseteq \pomo{1}$. Therefore, 
$(p_1,p) \in \beta$. By Lemma \ref{lem:greater-in-beta} and 
$p \leqlattice p_2$, we have $(p_1,p_2) \in \beta$.

\section{Transfer functions for lock/unlock instructions under \RA} 
\label{appendix:abs-sem}
\begin{figure}[!h]
	\centering
	\proofrule{unlock}
	{\begin{trgather} (pre(\lab),mo,m) \in \mathcal{S} \qquad \lab_l = \mathtt{FindLock}(\lab) \\ 
			\lab_l \in \mathtt{Lasts}(mo) \qquad mo'=mo[x \rightarrow mo(l) \append \lab] \end{trgather}}
	{\mathcal{S} \xrightarrow{\lab: \unlockinst{l}} \mathcal{S} \squplus (\lab, mo',m')}
	\\ \ \\ \ \\
	\proofrule{lock}
	{\begin{trgather} \sigma=\mathtt{PreProcessLock(\lab)} \qquad \lab_{ul} \in I(\lab) \\
			(\lab_{ul}, mo_{ul},m_{ul}) \in \mathcal{S} \qquad	(pre(\lab), mo, m) \in \sigma \\ 
			(pre(\lab), mo'', m')) = \mathtt{AI}((pre(\lab),mo , m),(\lab_{ul}, mo_{ul}, m_{ul})) \\
			mo' = mo'' \append \lab 
	\end{trgather}}
	{\mathcal{S} \xrightarrow{\lab: \lockinst{l}} \mathcal{S} \squplus (\lab, mo',m')}
	\caption{Transfer functions for Lock/Unlock instructions in \RA programs} \label{fig:lock-sem}
\end{figure}

The rules \textsc{lock} and \textsc{unlock} in figure \ref{fig:lock-sem} 
shows transfer functions 
for lock and unlock instructions over mutex variable $ l $ respectively. 
Whenever an unlock instructions is encountered, we check that the
corresponding lock instruction $ \lab_l $ should be at the end of partial order 
of mutex variable $ l $ in program states of location $ pre(\lab) $.
Intuitively, it means that the 
lock must be acquired by current thread in all program states
at program location $ pre(\lab) $.
We change the program state by appending 
the unlock instruction $ \lab $ in the partial order 
of mutex variable $ l $. 

The rule for lock instruction is a bit tricky. It uses following 
helper functions:
\begin{description}
	\item[\texttt{Lasts}($ p $):] Returns the set of last elements in a partial 
	order. Formally $ \{a \mid \nexists b \in \modord{Q}{x}\ .\ (a,b) \in \pomo{}\} $
	
	\item [\texttt{EndsInLock}($mo$):] True if the last of a 
	partial order contains 	some lock instruction, otherwise 
	false, i.e. it returns 
	$ \exists \lab_l \in \mathtt{Lasts}(mo(l)) \andspace \lab_l \in Locks(l) $
	
	\item [\texttt{FindUnlock}($\lab$)] Return the unlock 
	instruction corresponding 
	to lock instruction $ \lab $.
	
	\item [\texttt{FindLock}($\lab$)] Return the Lock 
	instruction corresponding 
	to unlock instruction $ \lab $.
	
	\item [\texttt{PreProcessLock}$(\lab)$] The function 
	first checks if in some program state at $ pre(\lab) $ 
	the mutex variable $ l $ is already acquired by 
	some thread. This can be done by checking if 
	partial order of mutex variable $ l $ ends with 
	a lock instruction $ \lab_l $. 
	If it ends in some other lock instruction $ \lab_l $, 
	\texttt{PreProcessLock} finds the unlock instructions 
	$ \lab_{ul} $ corresponding to $ \lab_l $ and apply 
	interference from this.	It combines the resulting 
	program states with the program state at $ pre(\lab) $
	that do not end in lock instruction. i.e. 
	$\mathtt{PreProcessLock}(\lab)$ returns 
	$\sigma$, where $\sigma \defeq  
	\forall (pre(\lab), mo, m) \in \mathcal{S}$ if $\mathtt{EndsInLock}(mo)$, then 
	$\lab_{ul} = \mathtt{FindUnlock}(\lab_l)$, $\forall (\lab_{ul}, mo_{ul}, m_{ul}) \in \mathcal{S}$,
	$\sigma = \sigma \squplus \mathtt{AI}((pre(\lab),mo , m),(\lab_{ul}, mo_{ul}, m_{ul}))$
	else $\sigma = \sigma \squplus (pre(\lab),mo , m)$
\end{description}

For lock instructions $\lab: \lockinst{l}$, we first perform the 
pre-processing step to get a list of program states in which no 
thread has acquired the lock over mutex variable $l$. Second, we apply the 
interference from all the unlock instructions over mutex 
variable $l$ in all the other threads to the current program state. 
This is required to make sure that we are considering all possible 
reorderings of locked regions. Finally, the lock 
instruction $ \lab $ is appended in the partial order corresponding to 
the mutex variable $ l $.


\end{appendix}

\end{document}